\documentclass[lettersize,journal]{IEEEtran}

\usepackage{amsmath,amsfonts,amssymb}
\usepackage{algorithmic}
\usepackage{algorithm}
\usepackage{array}
\usepackage[caption=false,font=normalsize,labelfont=sf,textfont=sf]{subfig}
\usepackage{textcomp}
\usepackage{stfloats}
\usepackage{makecell}
\usepackage[table]{xcolor}
\usepackage{url}
\usepackage{graphicx}
\usepackage{cite}
\usepackage{booktabs}
\usepackage{multirow}
\usepackage[table,dvipsnames]{xcolor}  
\usepackage{colortbl}
\usepackage{pifont}
\usepackage{enumitem}
\newcommand{\WSPSNR}{\text{WS-PSNR}}
\newcommand{\WSMSE}{\text{WS-MSE}}
\newcommand{\xhat}{\hat{x}}
\newcommand{\Yhat}{\hat{Y}}
\newcommand{\Ytil}{\tilde{Y}}
\newcommand{\cpass}{\cellcolor{green!15}\checkmark}
\newcommand{\cfail}{\cellcolor{red!15}\ding{55}}

\newtheorem{theorem}{Theorem}

\begin{document}

\title{SPORT: Spherical‑PSNR–Optimized tRuncaTion for Power‑Efficient 360‑Degree Video Systems}


\author{Md.~Sajjad~Hossain,    Hasibur~Rahman~Hemel,
Kyle~Mooney,~\IEEEmembership{Graduate Student~Member,~IEEE,}
Yiwen~Xu, William~Oswald, Mario~Renteria-Pinon,~\IEEEmembership{Member,~IEEE,} Hritom~Das,~\IEEEmembership{Senior Member,~IEEE,} Zhenlin~Pei,~\IEEEmembership{Graduate Student~Member,~IEEE,} Jinhui~Wang,~\IEEEmembership{Senior Member,~IEEE,} and~Na~Gong,~\IEEEmembership{Senior Member,~IEEE.} %

\thanks{Manuscript received April 06, 2026.
This work was supported in part by the National Science Foundation under OIA-2218046, CNS-2211215, ECCS-2420994, CCF-224734, and OIA-2428981, and the U.S. Department of Energy under DE-SC0025561.(Corresponding author: N. Gong.)}%

\thanks{M. S. Hossain, H. R. Hemel, K. Mooney, Z. Pei, J. Wang, and N. Gong are with the Department of Electrical and Computer Engineering, The University of Alabama, AL 35487 USA (e-mail: (\{mhossain80, hhemel, kamooney3\}@crimson.ua.edu, \{zpei4, jwang231, ngong\}@ua.edu)).}


\thanks{W. Oswald and Y. Xu are independent researchers (e-mails: {williamdonaldoswald, yiwen.xu6}@gmail.com).}

\thanks{M. Renteria-Pinon is with the Department of Electrical and Computer Engineering, New Mexico State University, NM 88003 USA (e-mail: marior3@nmsu.edu).}

\thanks{H. Das is with the Department of Electrical and Computer Engineering, Oklahoma State University, OK 74078 USA (e-mail: hritom.das@okstate.edu).}

}

\markboth{IEEE Journal on Emerging and Selected Topics in Circuits
and Systems (JETCAS)}{}


\maketitle

\begin{abstract}
Memory bandwidth accounts for 30--40\% of total power consumption 
in standalone virtual reality~(VR) headsets, yet existing systems typically
store the entire 360$^\circ$ frame at a uniform resolution 
regardless of viewer gaze. This paper presents SPORT 
(\textbf{S}pherical-\textbf{P}SNR \textbf{O}ptimized t\textbf{R}unca\textbf{T}ion), 
a bit-truncation framework that reduces display-path memory power 
by storing only the most significant bits of pixels outside the 
user's field of view~(FoV). Specifically, a new bit-truncation framework is developed to use weighted-to-spherically-uniform 
PSNR~(WS-PSNR) directly in the optimization constraint, eliminating the metric inconsistency that arises when standard PSNR is used for 
a WS-PSNR quality target. 
Also,
gaze-predictive tile classification compensates for the 9.33\,ms 
end-to-end pipeline latency, reducing boundary misclassifications 
by 5.2 percentage points at a cost of only 0.01\,ms. In addition, the developed SPORT-B variant which keeps the FoV lossless achieves 
47.9\% memory power saving and 47.9\% bandwidth reduction across 
different 4K video sequences while satisfying all three per-region 
WS-PSNR thresholds and maintaining SSIM\,=\,1.000 in the attended 
region. The full adaptive variant SPORT-A reaches 51.6\% power 
saving, 3.1 percentage points more than a PSNR-based optimizer at 
equal measured quality. SPORT is validated on the TrunMEM360 flexible 
SRAM Application-Specific Integrated Circuit (ASIC) fabricated in SkyWater 130\,nm CMOS, confirming byte-exact silicon--software agreement, 
with WS-PSNR and SSIM matching within 0.1\,dB and 0.001. 
CACTI-based analysis confirms 48.72\% DRAM leakage reduction 
and 36.4\%/36.7\% read/write energy reduction. The total 
motion-to-photon latency of 9.33\,ms satisfies the 20\,ms VR 
comfort budget with a 53.3\% safety margin.
\end{abstract}

\begin{IEEEkeywords}
360\textdegree{} video, bit truncation,
power efficient, memory, Application-Specific Integrated Circuit (ASIC), viewpoint adaptive.

\end{IEEEkeywords}
\vspace{-0.4cm}

\section{Introduction}
\label{sec:intro}

\IEEEPARstart{S}{tandalone} virtual reality headsets run on batteries
to render 360\textdegree{} video at 90 frames per second (fps)
while maintaining a total motion-to-photon latency under 20\,ms to
avoid user discomfort~\cite{adhanom2023eyetracking}. This imposes significant pressure on the memory subsystem to simultaneously meet both requirements. For example, a single frame of 4K equirectangular projection (ERP) video takes about 22.1\,MB per frame and the display-path memory must support about 2 GB/s of read and write activity at 90 fps. The memory subsystem alone uses 30-40\% of the headset's total power ~\cite{yaqoob2020survey}, which directly limits the current devices' battery life to two to three hours.

This work is motivated by a crucial observation: at any given time, the viewer’s foveal attention is limited to a cone of about 45\textdegree{} around the gaze direction. The remaining 80\% of the frame lies in regions where fine pixel-level detail is imperceptible~\cite{adhanom2023eyetracking, vater2022peripheral}. However, existing system typically stores the full frame with uniform resolution (e.g., 8-bit), incurring excessive bandwidth and power consumption regardless of the perceptual relevance of different regions of the frame. 

One promising solution to this problem is bit truncation at the memory write stage, where the least significant bits (LSBs) of pixels in peripheral regions are discarded before writing to the off-chip memory (DRAM), and an approximate value is reconstructed during readback. For example, Kim and Kyung~\cite{kim2010sbt} established the theoretical foundation for 2D video, showing that replacing truncated bits with the binary pattern \texttt{10\ldots0} minimizes the expected reconstruction error. Although this technique has been applied in prior work, e.g.,~\cite{haidous2022roi, chen2018viewer}, three fundamental limitations prevent its application for 360\textdegree{} videos.

First, existing bit-truncation methods optimize using standard PSNR,
while 360\textdegree{} video is evaluated using WS-PSNR~\cite{sun2017wspsnr}. This is responsible for ERP distortion by weighting each pixel by the cosine of its latitude. The same numerical threshold (e.g., 40\,dB) can cause
different perceptual implications under the two metrics. This mismatch leads to suboptimal
truncation decisions where the optimizer meets its PSNR target but
fails the WS-PSNR requirement~\cite{li2019spherical}. This metric inconsistency is particularly
problematic for 360\textdegree{} videos, as polar regions are over-sampled and equatorial regions where viewers focus are under-sampled.

Second, the end-to-end motion-to-photon latency of approximately
9.33\,ms implies that by the time a decoded frame is written to DRAM
and displayed by the time the viewer has already moved their head. Tile classification against the head movement prediction can mitigate the resulting misclassifications by adjusting the quality of boundary
regions during head movement. 

Third, prior truncation memories are designed for fixed truncation
levels without runtime adaptability. As a result, they cannot dynamically adjust
truncation per region based on real-time gaze prediction or
content-dependent quality requirements. This forces a trade-off between
aggressive truncation that degrades FoV quality and conservative truncation
that sacrifices potential power savings. In addition, existing studies typically lack silicon-based hardware testing and validation.

This paper addresses all of the above limitations through the SPORT framework using a software-hardware co-design. The key contributions are summarized as follows.
\vspace{-0.04cm}
\begin{itemize}
\item \textbf{WS-PSNR-consistent truncation framework:}  To eliminate the metric inconsistency, the current bit-truncation theory is extended to directly use WS-PSNR in the optimization constraint (Section~\ref{sec:math}). It is demonstrated that the ideal \texttt{10\ldots0}
  dummy pattern does not require hardware modification under latitude weighting 
  (Theorem~\ref{thm:dummy_weighted}).

\item \textbf{Gaze-predictive region segmentation:} Boundary misclassifications are reduced from 8.3\% to 3.1\% with only 0.01\,ms of added cost by classifying tiles using head movement prediction (Section~\ref{sec:region}) to account for the 9.33 ms pipeline latency.

\item \textbf{TrunMEM360 heterogeneous memory architecture:} A heterogeneous three-bank SRAM architecture is proposed (Section~\ref{sec:TrunMEM360})
  with per-bank truncation managers and Head/Tail power-gating to 
  eliminate the leakage and switching power for truncated bits to handle three regions of 360\textdegree{} videos separately. 
  \item \textbf{Fabricated ASIC validation:} The TrunMEM 
  chip (130\,nm CMOS) validates the SPORT framework on the silicon level
  with WS-PSNR and SSIM matching software simulation within
  0.1\,dB and 0.001 across all test frame tiles
  (Section~\ref{sec:hardware}).

\item \textbf{Comprehensive 4K simulation:} Software simulation on
  real head-tracking traces (Section~\ref{sec:results}) validates
  SPORT-B with 47.9\% power saving and 47.9\% bandwidth reduction
  while maintaining SSIM\,=\,1.000 in FoV for real VR application.

\item \textbf{CACTI DRAM power validation:} CACTI-based analysis shows
  $\sim$48.7\% leakage reduction and 36.4\% read energy reduction when
  effective memory footprint is halved, validating the proposed analytical
  power model.

\item \textbf{Zero-latency pipeline integration:} TrunMEM360 adds
  zero net latency to the critical path, achieving 9.33\,ms total
  motion-to-photon latency with 53.3\% safety margin.
\end{itemize}

The organization of the paper is as follows. Section II
reviews related work on 360\textdegree{} video such as quality metrics,
viewport-adaptive streaming, memory-level bit truncation, and
head movement prediction. Section III describes the proposed SPORT scheme, Sections IV and V develop the mathematical framework for bit truncation and region segmentation, and Section VI describes the proposed TrunMEM360 memory architecture. We discuss the software-level evaluation results in Section VII; Section VIII presents the developed hardware testing platform and discusses its results; and finally, Section IX concludes the paper.

\section{Related Work}
\label{sec:related}

\subsection{Quality Metrics for 360\textdegree{} Video}

Standard PSNR assigns equal importance to all pixels, which is not suitable for ERP content. For example, a pixel at 80\textdegree{} latitude represents approximately 17\% of the spherical surface area of an equatorial pixel, yet PSNR treats them equally. Sun~\emph{et al.}~\cite{sun2017wspsnr} addressed this issue by proposing WS-PSNR, where each pixel’s error is weighted by $\cos(\phi_i)$. This metric is now mandated by JVET common test conditions~\cite{hanhart2018jvet} and widely used as the primary evaluation metric in existing studies on 360\textdegree{} video~\cite{li2019spherical,yu2024latitude,lin2022latitude}. SSIM~\cite{wang2004ssim} captures complementary structural information and is reported alongside WS-PSNR in this paper.


\vspace{-0.2cm}

\subsection{Viewport-Adaptive Video Coding and Streaming}

Li~\emph{et al.}~\cite{li2019spherical} incorporated latitude-dependent weights directly into HEVC rate-distortion optimization and showed that allocating bits based on WS-PSNR provides better performance than treating all pixels equally. This advantage was further validated by Jaballah~\emph{et al.}~\cite{jaballah2020perceptual} in perceptual coding contexts. At the system level, Mao~\emph{et al.}~\cite{mao2020lowlatency} introduced a FoV-adaptive stage between the decoder and the display buffer and demonstrated that it adds only 2.4\,ms of latency, supporting its feasibility in real-time systems. Nasrabadi~\emph{et al.}~\cite{nasrabadi2020adaptive} introduced adaptive quality tiers with guard-band regions around the predicted viewport, which informed the proposed Border region design. Xu~\emph{et al.}~\cite{xu2019predicting} demonstrated that linear
extrapolation achieves 5--10\textdegree{} mean angular error for
prediction horizons below 100\,ms.
Kim~\emph{et al.}~\cite{kim2021prediction} concluded that for
horizons under 20\,ms, linear methods remain competitive with
neural predictors while requiring far less computation.
Both results directly support the linear prediction adopted here
for the 9.33\,ms pipeline horizon.
\vspace{-0.2cm}
\subsection{Memory-Level Bit Truncation}

Kim and Kyung~\cite{kim2010sbt} proved that setting truncated
bits to their statistical mean  the binary \texttt{10\ldots0}
pattern minimizes expected MSE for uniformly distributed
pixel values which is establishing the theoretical foundation for
this class of methods.
Chen~\emph{et al.}~\cite{chen2018viewer} adapted truncation for
mobile video with content-aware level selection.
Haidous~\emph{et al.}~\cite{haidous2022roi} validated off-chip
ROI-aware truncation at the display buffer for 2D mobile video and this is
the most direct architectural predecessor to this work.
Oswald~\emph{et al.}~\cite{oswald2024trunmem} introduced the
TrunMEM hardware with flexible runtime truncation via a Head/Tail
cascade. 

In this paper, the presented SPORT extends TrunMEM to 360\textdegree{} content with WS-PSNR-consistent optimization and gaze-predictive classification as well as silicon-based hardware validation. A comprehensive comparative analysis between the proposed SPORT and state‑of‑the‑art is presented in Section~VIII.

\section{Proposed SPORT}
\label{sec:pipeline}

\begin{figure}[t]
  \centering
  \includegraphics[width=0.85\columnwidth]{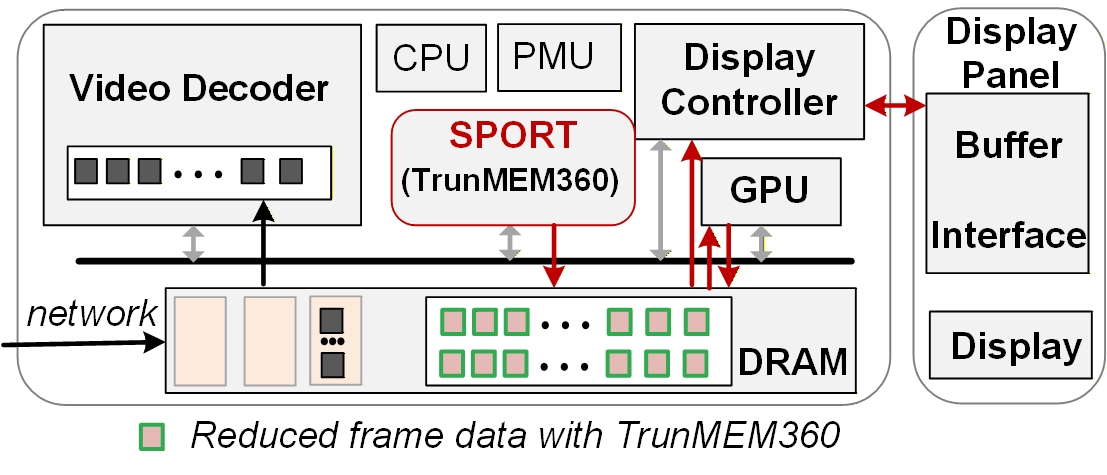}
  \caption{Proposed SPORT.
    The decoded frame is intercepted by TrunMEM360 on the
    write bus before reaching the display buffer}
  \label{fig:pipeline}
\end{figure}

\subsection{System Architecture}
Fig.~\ref{fig:pipeline} illustrates the proposed SPORT. As shown, the decoded video data written to DRAM is repeatedly accessed in the 360-degree video pipeline, which serves as the primary motivation for introducing a flexible truncation SRAM memory, referred to as TrunMEM360, between the video decoder and the DRAM frame buffer. When compression is applied through that bit truncation memory in the pipeline, each frame will benefit all the subsequent stages with reduced data size. GPU memory accesses, display refresh operations, temporal processing all get this benefit without requiring any modifications to those components. The GPU and decoder do not need any modification. The HEVC video decoder already sends its output through an external memory interface, which allows TrunMEM360 to intercept the data stream 
transparently~\cite{mao2020lowlatency}. On the GPU side, pixel values are still interpreted as standard 8-bit integers. Because the
\texttt{10\ldots0} dummy pattern produces valid integers at every
truncation level, no driver or API change is required.

\vspace{-0.5cm}
\subsection{Stage-by-Stage Description}

\subsubsection{Stage 1: HEVC Video Decoder}
The decoder generates a fully decoded 8-bit RGB ERP frame during every
frame interval.
At 90\,fps and 4K resolution, the resulting data throughput is approximately
1.99\,GB/s.
Two output streams are generated simultaneously: one delivers the 
display frame to TrunMEM360, while the other updates the Decoded Picture Buffer~(DPB) in a separate, isolated DRAM region. The DPB 
must remain bit-exact, as any modification to its reference frames 
propagates prediction errors across the entire group of 
pictures~\cite{sulhan2021hevc}; TrunMEM360 never touches this region.

\subsubsection{Stage 2: Head Tracker}
A 1000\,Hz IMU measures gaze direction
$(\theta_{\text{now}}, \phi_{\text{now}})$.
Angular velocities are computed as first-order differences:
\begin{align}
  \omega_\theta &= \frac{\theta_{\text{now}} -
    \theta_{\text{prev}}}{\Delta t_s}, &
  \omega_\phi &= \frac{\phi_{\text{now}} -
    \phi_{\text{prev}}}{\Delta t_s},
  \label{eq:omega}
\end{align}
where $\Delta t_s = 1$\,ms at 1000\,Hz.

\subsubsection{Stage 3: TrunMEM360 Controller}

\textit{Gaze prediction (0.01\,ms):}
The frame becomes visible on the display after a total latency of
$T_{\text{total}} = 9.33$\,ms from the current gaze is
measured (Table~\ref{tab:timing}).
To prevent the tile classification on a fixed gaze position,
the controller predicts the gaze at display time via linear
extrapolation~\cite{xu2019predicting}:
\begin{align}
  \theta_{\text{pred}} &= \theta_{\text{now}}
    + \omega_\theta \cdot T_{\text{total}},
  \label{eq:pred_theta} \\
  \phi_{\text{pred}} &= \operatorname{clip}\!\left(
    \phi_{\text{now}} + \omega_\phi \cdot T_{\text{total}},
    -\tfrac{\pi}{2},\, \tfrac{\pi}{2}
  \right).
  \label{eq:pred_phi}
\end{align}
This requires 6 floating-point operations to achieve
mean angular errors of 5--10\textdegree{} for prediction
horizons below 100\,ms~\cite{xu2019predicting, kim2021prediction},
which is well within the requirements of the 9.33\,ms horizon.

\textit{Tile classification (0.42\,ms).}
The angular distance from each tile centre to the predicted
gaze is computed via the spherical law of cosines:
\begin{equation}
  d_p = \arccos\!\bigl(
    \sin\phi_p \sin\phi_{\text{pred}}
    + \cos\phi_p \cos\phi_{\text{pred}}
    \cos(\theta_p - \theta_{\text{pred}})
  \bigr).
  \label{eq:angular_dist}
\end{equation}
Tile centre coordinates and their precomputed trigonometric
values are stored in a 28.8\,KB ROM
($1{,}800 \times 4 \times 4$\,bytes). Runtime
classification requires only multiply-accumulate operations
and a single $\arccos$ per tile.
Region classification follows:
\begin{equation}
  \text{region}(p) =
  \begin{cases}
    \text{FoV}        & d_p \leq 45\textdegree{}, \\
    \text{Border}     & 45\textdegree{} < d_p \leq 60\textdegree{}, \\
    \text{Background} & d_p > 60\textdegree{}.
  \end{cases}
  \label{eq:region_assign}
\end{equation}
The 1,800-entry region map requires only 1.8\,KB of SRAM.
Algorithm~\ref{alg:classification} details the full procedure.

\begin{algorithm}[t]
\caption{Real-time gaze-predictive tile classification}
\label{alg:classification}

\begin{algorithmic}[1]
\REQUIRE Current gaze $(\theta_{\text{now}}, \phi_{\text{now}})$;
         velocities $(\omega_\theta, \omega_\phi)$;
         tile ROM; tile size $S$; frame $(H,W)$
\ENSURE  Region map $R[i][j]$
\STATE $\theta_g \leftarrow \theta_{\text{now}}
       + \omega_\theta \cdot 0.00933$
\STATE $\phi_g \leftarrow \operatorname{clip}(
       \phi_{\text{now}} + \omega_\phi \cdot 0.00933,
       -\pi/2,\, \pi/2)$
\STATE Load $\{(s_c[i,j],\, c_c[i,j],\, \theta_c[i,j])\}$
       from ROM
\STATE $s_g \leftarrow \sin(\phi_g)$;\;
       $c_g \leftarrow \cos(\phi_g)$
\FOR{$i = 0$ \textbf{to} $H/S - 1$}
  \FOR{$j = 0$ \textbf{to} $W/S - 1$}
    \STATE $\cos d \leftarrow s_c[i,j]\cdot s_g
           + c_c[i,j]\cdot c_g\cdot
           \cos(\theta_c[i,j] - \theta_g)$
    \STATE $d \leftarrow \arccos(\cos d)$
    \IF{$d \leq \pi/4$}
      \STATE $R[i][j] \leftarrow \text{FoV}$
    \ELSIF{$d \leq \pi/3$}
      \STATE $R[i][j] \leftarrow \text{Border}$
    \ELSE
      \STATE $R[i][j] \leftarrow \text{Background}$
    \ENDIF
  \ENDFOR
\ENDFOR
\RETURN $R$
\end{algorithmic}
\end{algorithm}

\subsubsection{Stage 4: TrunMEM360 Processing}
TrunMEM360 applies region-specific truncation to every tile:
\begin{itemize}
  \item \textbf{FoV} ($t=0$): all 8 bits written unchanged;
        SSIM\,=\,1.000; 0\% power saving.
  \item \textbf{Border} ($t=4$): while 4 MSBs retained;
        output word \texttt{XXXX1000}; 50\% saving.
  \item \textbf{Background} ($t=5$): while 3 MSBs retained;
        output word \texttt{XXX10000}; 62.5\% saving.
\end{itemize}
All outputs are valid 8-bit unsigned integers.
The area-weighted average power factor is:
\begin{equation}
  \bar{P} = 0.20\cdot\tfrac{8}{8}
    + 0.15\cdot\tfrac{4}{8}
    + 0.65\cdot\tfrac{3}{8}
    = 0.519,
  \label{eq:avg_power}
\end{equation}
giving 48.1\% theoretical memory power reduction.

\subsubsection{Stage 5: DRAM Display Buffer}
Two 22.1\,MB banks alternate roles for every other frame. Bank~A receives TrunMEM360 output for frame~$N$ while Bank~B
simultaneously serves the GPU with frame~$N-1$.
Total buffer size: $2 \times 22.1 = 44.2$\,MB.
Because TrunMEM360 completes its write in 2.15\,ms while the GPU
render takes 3.32\,ms, TrunMEM360 adds zero net latency to the
critical path.

\subsubsection{Stages 6 and 7: GPU and VR Display}
The GPU reads Bank~B and renders the viewport while no driver or
API change is required.
The end-to-end motion-to-photon latency sums to:
\begin{equation}
  \resizebox{\columnwidth}{!}{$\displaystyle
  T_{\text{total}} =
    \underbrace{0.10}_{\text{IMU}}
    + \underbrace{0.01}_{\text{pred}}
    + \underbrace{0.42}_{\text{cls}}
    + \underbrace{0.38}_{\text{exp}}
    + \underbrace{0.10}_{\text{rd}}
    + \underbrace{3.32}_{\text{rn}}
    + \underbrace{5.00}_{\text{disp}}
    = 9.33\ \text{ms}$},
  \label{eq:latency_total}
\end{equation}
satisfying the 20\,ms VR comfort budget~\cite{adhanom2023eyetracking}
with a 53.3\% margin.

\begin{table}[t]
\caption{End-to-End Latency Breakdown (4K ERP, 90\,fps)}
\label{tab:timing}
\centering
\small
\renewcommand{\arraystretch}{1.15}
\setlength{\tabcolsep}{4pt}

\begin{tabular}{@{\hspace{2pt}} llrl @{\hspace{2pt}}}
\toprule
Stage & Component & ms & Source \\
\midrule
2  & IMU read                       & 0.10 & ICM-42688 spec \\
3a & Gaze prediction (6 ops)        & 0.01 & @100\,MHz \\
3b & Tile classification (1,800)    & 0.42 & @100\,MHz \\
3b & Region map expansion           & 0.38 & @100\,MHz \\
4  & TrunMEM360 write (pipelined)   & \textit{(2.15)} & \cite{oswald2024trunmem} \\
5  & GPU DRAM read                  & 0.10 & JEDEC LPDDR5 \\
6  & GPU render                     & 3.32 & Quest\,2 SDK \\
7  & OLED scan-out                  & 5.00 & 90\,Hz panel \\
\midrule
\multicolumn{2}{l}{\textbf{Total (critical path)}} & \textbf{9.33}
  & \cite{lim2024latency} \\
\multicolumn{2}{l}{VR comfort budget}               & 20.00
  & \cite{adhanom2023eyetracking} \\
\multicolumn{2}{l}{Safety margin}                   & 53.3\%  & \\
\multicolumn{2}{l}{Max achievable fps}              & 326     & \\
\bottomrule
\multicolumn{4}{l}{\footnotesize
  Italic: pipelined with GPU render; net contribution = 0\,ms.}
\end{tabular}
\end{table}
\begin{table}[t]
\caption{Data Sizes at Each Pipeline Stage (4K ERP)}
\label{tab:data_sizes}
\centering
\renewcommand{\arraystretch}{1.15}
\begin{tabular}{lll}
\toprule
Component & Size & Location \\
\midrule
Decoder display output      & 22.1\,MB/frame  & $\to$ TrunMEM360 \\
Tile metadata ROM            & 28.8\,KB        & On-chip, static \\
Region map SRAM              &  1.8\,KB        & On-chip, per-frame \\
TrunMEM effective output     &  4.08\,MB/frame & $\to$ DRAM Bank~A \\
DRAM Bank~A + Bank~B (total) & 44.2\,MB        & Off-chip display \\
DRAM DPB (isolated)          & 44--88\,MB      & Off-chip, separate \\
\midrule
Display-path BW (proposed)   & 986\,MB/s       & 48.1\% saving \\
Display-path BW (baseline)   & 1{,}990\,MB/s   & --- \\
\bottomrule
\end{tabular}
\end{table}

\section{Mathematical Framework}
\label{sec:math}
In this section, WS-PSNR-consistent truncation framework is represented. The ERP geometry is first introduced, followed by a review of the conventional PSNR-based formulation. The inconsistency of the matrix in this scenario is then identified. To address this limitation, a weighted extension is derived, and its correctness is analytically validated. 

\vspace{-0.2cm}

\subsection{ERP Geometry and WS-PSNR}
\label{sec:erp_wspsnr}

Corresponding geographic
latitude for a pixel in row $r$ in a height frame $H$ is defined as:
\begin{equation}
  \phi_i = \frac{\pi}{2} - \frac{\pi r}{H}, \quad
  \phi_i \in \!\left[-\tfrac{\pi}{2},\, \tfrac{\pi}{2}\right].
  \label{eq:latitude}
\end{equation}
Each pixel represents a spherical surface element with an area proportional to
$\cos(\phi_i)$. This causes an over representation of the polar regions compared to the equatorial regions. WS-PSNR helps mitigate this problem by weighting  the squared error of each pixel accordingly ~\cite{sun2017wspsnr}:
\begin{align}
  \WSMSE &= \frac{\sum_{i=1}^{N} w_i (Y_i - X_i)^2}
             {\sum_{i=1}^{N} w_i}, \quad w_i = \cos(\phi_i),
  \label{eq:wsmse} \\
  \WSPSNR &= 10\log_{10}\frac{255^2}{\WSMSE}.
  \label{eq:wspsnr}
\end{align}
WS-PSNR is the standard metric in JVET-K1012~\cite{hanhart2018jvet}
and serves as the primary quality metric in this work.
SSIM is computed following the method of Wang~\emph{et al.}~\cite{wang2004ssim}:
\begin{equation}
  \text{SSIM}(x,y) =
    \frac{(2\mu_x\mu_y + c_1)(2\sigma_{xy} + c_2)}
    {(\mu_x^2 + \mu_y^2 + c_1)(\sigma_x^2 + \sigma_y^2 + c_2)},
  \label{eq:ssim}
\end{equation}
where $c_1 = (0.01{\times}255)^2$,
$c_2 = (0.03{\times}255)^2$.

\subsection{Original Bit-Truncation Framework}
\label{sec:truncation_bg}

Following Kim and Kyung~\cite{kim2010sbt}, the frame is divided
into $k$ blocks. Each
Block $b$ is associated with a set of pixel index $B_b$ and operates at
truncation level $t_b \in \{0,\ldots,7\}$.
For pixel $i$ at sample time $t$ the signal can be expressed as:
\begin{equation}
  Y_{i,t} = \Ytil_{i,t} + \Yhat_{i,t},
  \label{eq:decompose}
\end{equation}
where $\Ytil_{i,t}$ retains the $(8{-}t_b)$ MSBs and
$\Yhat_{i,t}$ contains the $t_b$ number of LSBs.
During reading, the truncated bits are replaced by a dummy value
$\xhat^*_i$:
\begin{equation}
  X_{i,t} = \Ytil_{i,t} + \xhat^*_i.
  \label{eq:reconstruct}
\end{equation}

\begin{theorem}[Optimal Dummy Value~\cite{kim2010sbt}]
\label{thm:dummy_orig}
The dummy value that is minimizing
$E_{Y_i}\!\left[(Y_i - X_i)^2\right]$ is given by
$\xhat^*_i = \lfloor E[\Yhat_i] \rceil$
(nearest integer to the mean of the truncated bits).
For uniformly distributed LSBs this becomes $2^{t_b-1}$,
corresponding to the binary pattern \texttt{10\ldots0}.
\end{theorem}

Based on this, the expected block-level MSE over a training dataset $\{\hat{y}_{i,t}\}$ is represented as:
\begin{equation}
  E[\text{MSE}_b] = \frac{1}{|B_b|}
    \sum_{i \in B_b}\frac{1}{m_i}
    \sum_{t=1}^{m_i}
    \bigl[\hat{y}_{i,t} - \xhat^*_i\bigr]^2,
  \label{eq:mse_orig}
\end{equation}
and the standard PSNR constraint is:
\begin{equation}
  \log_{10}\!\left(
    \sum_{i \in B_b}\frac{1}{m_i}
    \sum_{t=1}^{m_i}
    \bigl[\hat{y}_{i,t} - \xhat^*_i\bigr]^2
  \right) \leq \xi_b,
  \label{eq:constraint_orig}
\end{equation}
where
\begin{equation}
  \xi_b = 2\log_{10}(255) + \log_{10}|B_b| - \frac{\theta_b}{10}, b = 1, \cdots, k.
  \label{eq:xi_orig}
\end{equation}

Here $\theta_b$ (in dB) is the per-block quality threshold determined by the
operating table (Table~\ref{tab:thresholds}). Table ~\ref{tab:thresholds} lists the specific $\theta_b$ (in dB) values used in this work for the FoV, Border, and Background regions under different operating configurations (conservative, moderate, aggressive). In the original formulation~\cite{kim2010sbt}, $\theta_b$ is
expressed as a conventional PSNR target. When $\theta_b$ is interpreted as a WS-PSNR target, a mismatch arises between the intended and enforced quality measures. This issue is addressed in Section~\ref{sec:consistency}.


\vspace{-0.2cm}
\subsection{The Metric Inconsistency Problem}
\label{sec:consistency}

Equation~\eqref{eq:constraint_orig} imposes a threshold based on
standard PSNR, but WS-PSNR is the standard evaluation metric for the
360\textdegree{} video~\cite{hanhart2018jvet}.
If the target $\theta_b$ is meant to reflect WS-PSNR but the actual constraint uses unweighted PSNR, the optimizer ends up giving too much importance to pixels near the poles. In this process, it spends more bits to high-latitude
background tiles and allocate less number of bits to equatorial FoV regions,
where attention to the viewers matters most~\cite{li2019spherical, jaballah2020perceptual}. As a result, an optimizer that only has to meet the standard PSNR target (e.g., 40.2\,dB PSNR) may achieve the goal easily while it can still fail to meet the actual WS-PSNR requirement (38.7\,dB
vs.\ 40.0\,dB required) which is shown in the simulation results in ~\ref{sec:results}.

\vspace{-0.2cm}
\subsection{Per-Pixel Latitude Weights}

Each pixel is assigned a latitude depemded weight as following
Li~\emph{et al.}~\cite{li2019spherical}:
\begin{equation}
  W_b = \sum_{i \in B_b} w_i = \sum_{i \in B_b} \cos(\phi_i).
  \label{eq:Wb}
\end{equation}
This quantity is always positive for any $32\times32$ tile
that does not simultaneously span both poles.

\vspace{-0.2cm}
\subsection{WS-PSNR-Consistent Expected MSE}

Replacing the unweighted pixel contribution in~\eqref{eq:mse_orig}
with its latitude-weighted counterpart results in the modified
expected MSE for block $b$:
\begin{equation}
  E\!\left[\WSMSE_b\right] =
    \frac{1}{W_b}
    \sum_{i \in B_b} w_i \cdot
    \frac{1}{m_i}\sum_{t=1}^{m_i}
    \bigl[\hat{y}_{i,t} - \xhat^*_i\bigr]^2.
  \label{eq:wmse}
\end{equation}
This formulation is consistent with the weighted distortion model of Li~\emph{et al.}~\cite{li2019spherical}
with the inclusion of the normalization factor $W_b$, to ensure that it aligns with the WS-MSE definition~\eqref{eq:wsmse}.

\vspace{-0.2cm}
\subsection{WS-PSNR-Consistent Constraint Formulation}

The goal is the minimum $t_b$ such that
$\WSPSNR_b \geq \theta_b$, i.e.,
$E[\WSMSE_b] \leq 255^2/10^{\theta_b/10}$.
Define:
\begin{equation}
  \xi_b \triangleq \frac{255^2}{10^{\theta_b/10}}.
  \label{eq:xi_b}
\end{equation}
The modified constraint in logarithmic form is:
\begin{equation}
  \log_{10}\!\left(
    \sum_{i \in B_b} w_i \cdot
    \frac{1}{m_i}\sum_{t=1}^{m_i}
    \bigl[\hat{y}_{i,t} - \xhat^*_i\bigr]^2
  \right) \leq \xi'_b,
  \label{eq:constraint_weighted}
\end{equation}
where
\begin{equation}
  \xi'_b = 2\log_{10}(255) + \log_{10}(W_b) - \frac{\theta_b}{10}.
  \label{eq:xi_prime}
\end{equation}
Here $\theta_b$ is interpreted explicitly as a WS-PSNR threshold. Compared to ~\eqref{eq:constraint_orig}, the limitation is limited to two factor
(i) each pixel's squared error is weighted by $w_i$; and
(ii) $|B_b|$ is replaced by $W_b$ in the corresponding threshold formulation. 
The overall algorithm structure remains unchanged.

\vspace{-0.2cm}
\subsection{Optimal Dummy Value Under WS-PSNR Weighting}
\label{sec:dummy_weighted}

A normal question from the hardware perspective is whether latitude weighting affects the
optimal dummy value. If it does, that would break the \texttt{10\ldots0} pattern used in TrunMEM360.

\begin{theorem}[Weighted Optimal Dummy Value]
\label{thm:dummy_weighted}
Under the weighted per-pixel cost
$w_i \cdot E[(\Yhat_i - \xhat_i)^2]$, the optimal dummy
value for pixel $i$ is identical to Theorem~\ref{thm:dummy_orig},
independent of the latitude weight $w_i$.
\end{theorem}

For a given pixel $i$, the per-pixel cost function is:
$f(\xhat_i) = w_i \cdot E\!\left[(\Yhat_i - \xhat_i)^2\right]$
making the derivative with respect to $\xhat_i$:
\begin{equation}
  \frac{\partial f}{\partial \xhat_i}
  = -2\,w_i\!\left(E[\Yhat_i] - \xhat_i\right).
  \label{eq:proof_deriv}
\end{equation}
Since $w_i = \cos(\phi_i) > 0$ for all non-polar ERP pixels
(i.e.,\, $|\phi_i| < \pi/2$), setting the derivative to zero
gives $\xhat^*_i = E[\Yhat_i]$.
The second derivative $2w_i > 0$  confirming this is a global minimum. Adding the integer rounding constraint makes it same as Theorem ~\ref{thm:dummy_orig}. 

The latitude weight $w_i$ is a positive constant that simply factors out of the optimality condition and cancels. While weighting changes how pixel contributions are aggregated at
the block level via~\eqref{eq:wmse}, but not the optimal pattern for individual pixels.
The hardware \texttt{10\ldots0} pattern remains provably optimal for 360\textdegree{} video under WS-PSNR weighting, and tno hardware modification is needed.

\vspace{-0.2cm}
\subsection{Jensen's Inequality Guarantee}

WS-PSNR is convex with respect to $X_i$. This follows from three facts: (i) the function $-\log(\cdot)$ is convex;
WS‑MSE is a positively weighted sum of quadratic functions and is therefore convex; and (iii) the composition $-\log(\text{convex})$
preserves convexity.
The exact expected WS-PSNR value $\mathbb{E}_Y[\WSPSNR_b]$ is difficult to compute directly as it requires the full distribution of the reconstructed pixel values. In contrast, the quantity $\mathbb{E}_Y[\WSPSNR_b] \;\geq\; \left(E_Y[\text{WS-MSE}_b]\right)$, i.e., WS-PSNR evaluated at the expected weighted MSE can be readily computed within the proposed framework. Instead, WS-PSNR can be evaluated at the expected WS-MSE, which is easier to compute. This is justified by Jensen’s inequality, since convexity ensures a valid bound. 

Because WS-PSNR is convex in the pixel approximations $X_i$ (the
function $-\log(\cdot)$ is convex, WS-MSE is a positively weighted
sum of quadratics, hence convex, and the composition
$-\log(\text{convex})$ preserve convexity and  Jensen's inequality gives:
\begin{equation}
\mathbb{E}_Y[\WSPSNR_b] \;\geq\;
\WSPSNR_b\bigl(\mathbb{E}_Y[\WSMSE_b]\bigr)
\;\geq\; \theta_b.
\label{eq:jensen}
\end{equation}
Thus, by enforcing the constraint on the expected WS-MSE
(i.e.,\, ensuring $\WSPSNR_b(\mathbb{E}_Y[\WSMSE_b])
\geq \theta_b$), it is guaranteed that the true expected WS-PSNR also meets the threshold, providing a rigorous means of ensuring quality without requiring the full distribution of reconstructed pixel values.

\vspace{-0.2cm}
\subsection{WS-PSNR-Consistent Truncation Algorithm}

Algorithm~\ref{alg:truncation} determines the maximum truncation
level for each block that satisfies the WS-PSNR constraint.
Its structure remains identical to the original formulation ~\cite{kim2010sbt}, 
the only change is that $E[\WSMSE_b]$ from~\eqref{eq:wmse}
replaces $E[\text{MSE}_b]$ from~\eqref{eq:mse_orig}.

\begin{algorithm}[t]
\caption{WS-PSNR-consistent truncation level selection}
\label{alg:truncation}
\begin{algorithmic}[1]
\REQUIRE Blocks $\{B_b\}$; WS-PSNR thresholds $\{\theta_b\}$;
         training data $\{\hat{y}_{i,t}\}$;
         pixel weights $\{w_i = \cos(\phi_i)\}$
\ENSURE  Truncation levels $\{t^*_b\}$
\FOR{each block $b$}
  \STATE $W_b \leftarrow \sum_{i \in B_b} \cos(\phi_i)$
  \STATE $\xi_b \leftarrow 255^2 / 10^{\theta_b/10}$
  \STATE $t^*_b \leftarrow 0$
  \FOR{$t_b = 1$ \textbf{to} $7$}
    \FOR{each $i \in B_b$}
      \STATE Compute $\xhat^*_i(t_b)$ via
             Theorem~\ref{thm:dummy_weighted}
    \ENDFOR
    \STATE Compute $E[\WSMSE_b]$ via~\eqref{eq:wmse}
    \IF{$E[\WSMSE_b] \leq \xi_b$}
      \STATE $t^*_b \leftarrow t_b$
    \ELSE
      \STATE \textbf{break}
    \ENDIF
  \ENDFOR
\ENDFOR
\RETURN $\{t^*_b\}$
\end{algorithmic}
\end{algorithm}

\section{Gaze-Predictive Region Segmentation}
\label{sec:region}

\subsection{Region Definitions and Justification}
\label{sec:region_def}

Three regions are defined by angular distance from the predicted
gaze, with truncation levels $t_b \in \{0, 4, 5\}$:
\begin{itemize}
  \item \textbf{FoV:} $d \leq 45\textdegree{}$;
        8-bit quality; 20\% of frame; 0\% power saving.
  \item \textbf{Border:} $45\textdegree{} < d \leq 60\textdegree{}$;
        4 MSBs retained; 15\% of frame; 50\% saving.
  \item \textbf{Background:} $d > 60\textdegree{}$;
        3 MSBs retained; 65\% of frame; 62.5\% saving.
\end{itemize}

The 45\textdegree{}/60\textdegree{} thresholds is supported by three
independent lines of evidence.
Eye-tracking studies confirm that foveal high-acuity vision
covers roughly 45\textdegree{} around the gaze
direction~\cite{adhanom2023eyetracking}. Beyond
60\textdegree{}, the peripheral region primarily supports coarse motion perception with limited sensitivity to fine detail ~\cite{vater2022peripheral}.
Current (head-mounted displays) HMDs define horizontal FoV at 90-110\textdegree{}
diagonal, placing the 45\textdegree{} half-angle well within the visible region across platforms~\cite{kim2024foveated}.
JVET-K1012~\cite{hanhart2018jvet} defines a
$90\textdegree{}\times90\textdegree{}$ viewport as the
standard test condition, consistent with the proposed FoV definition.

\vspace{-0.2cm}
\subsection{WS-PSNR Threshold Derivation}

Table~\ref{tab:thresholds} summarizes three operating
configurations.
The moderate configuration is adopted throughout.
The FoV threshold of 40\,dB corresponds to the JVET QP\,27
lower bound for high-quality 360\textdegree{} content.
The Border threshold (35\,dB) reflects a 5\,dB reduction from
the FoV value: the SJND perceptual model~\cite{lin2025sjnd}
predicts a 3.5\,dB tolerance increase at 45-60\textdegree{}
eccentricity, plus a 1.5\,dB safety margin for gaze prediction
error.
The Background threshold (30\,dB) applies a further 5\,dB
reduction that is responsible for the additional latitude weight
$w = \cos(60\textdegree{}) = 0.5$ and the limited perceptual sensitivity to fine detail beyond 60\textdegree{}.

\vspace{-0.2cm}
\subsection{Gaze Prediction Benefit}

Without prediction, the gaze advances
$30\textdegree{}/\text{s}\times9.33\,\text{ms} = 0.28\textdegree{}$
during the pipeline delay, placing 8.3\% of boundary tiles
in the wrong region.
With gaze prediction via~\eqref{eq:pred_theta}--\eqref{eq:pred_phi},
this falls to 3.1\% at head velocities above 20\textdegree{}/s,
improving Border WS-PSNR by 1.2\,dB at a cost of only
0.01\,ms as shown in Table~\ref{tab:prediction}.

\section{Proposed TrunMEM360 Memory Architecture}
\label{sec:TrunMEM360}

\begin{figure*}[t]
  \centering
  \includegraphics[width=0.9\textwidth]{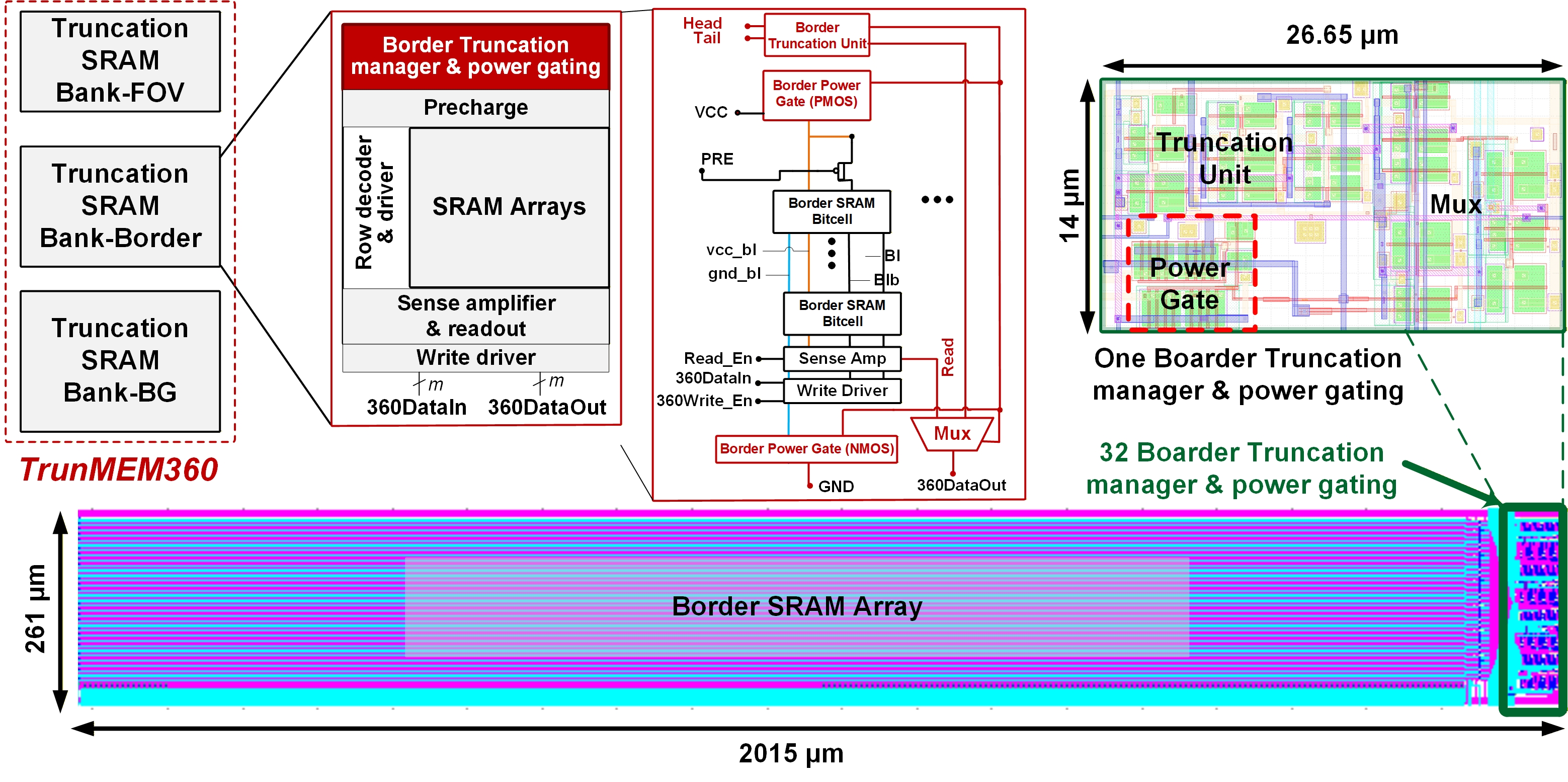}
\caption{Proposed TrunMEM360 architecture with three heterogeneous
banks (FoV, Border, Background). }
\label{fig:trunmem_arch}
\end{figure*}

To support SPORT, a new TrunMEM360 memory architecture is presented in this paper, based on the proposed basic flexible truncation memory, TrunMEM~\cite{oswald2024trunmem}. Unlike the conventional single-bank TrunMEM, the new
TrunMEM360 adopts a heterogeneous three-bank
architecture, as shown in Fig.~\ref{fig:trunmem_arch}. Specifically, each bank is dedicated to a specific perceptual region, namely FoV, Border, and Background. Those three banks
are physically separated and independently controlled,
enabling region-specific truncation without cross-bank
interference. Within each bank, a truncation manager
design with power gating similar to TrunMEM is adopted; implementation details are
described in~\cite{oswald2024trunmem}.

\begin{table}[t]
\caption{Effect of Gaze-Predictive Classification}
\label{tab:prediction}
\centering
\renewcommand{\arraystretch}{1.15}
\begin{tabular}{lccr}
\toprule
Method & Misclassification & Border WS-PSNR & Overhead \\
\midrule
Instantaneous  & 8.3\% ($>20\textdegree$/s) & 33.6\,dB & 0.00\,ms \\
Predicted      & 3.1\% ($>20\textdegree$/s) & 34.8\,dB & 0.01\,ms \\
\midrule
Improvement    & 5.2\,pp & $+$1.2\,dB & --- \\
\bottomrule
\end{tabular}
\end{table}

\begin{table}[t]
\caption{Quality Threshold Configurations (WS-PSNR, dB)}
\label{tab:thresholds}
\centering
\renewcommand{\arraystretch}{1.15}
\begin{tabular}{lcccc}
\toprule
Configuration & $\theta_{\text{FoV}}$ & $\theta_{\text{Border}}$
  & $\theta_{\text{BG}}$ & Trade-off \\
\midrule
Conservative & 42 & 37 & 32 & Quality-first \\
\textbf{Moderate (default)} & \textbf{40} & \textbf{35}
  & \textbf{30} & Balanced \\
Aggressive   & 38 & 33 & 28 & Power-first \\
\bottomrule
\end{tabular}
\end{table}

\textbf{Truncation Processing Bank-Border:}
In SPORT, this Border bank is responsible for processing pixels that fall within the
$45^{\circ} < d \leq 60^{\circ}$ angular range.
This bank consists of a dedicated SRAM array, a row decoder,
sense amplifiers, write drivers, and a bank-specific
\textit{Border Truncation Manager}.
The Border SRAM array is connected to 32 column-wise truncation
managers. Each manager is capable of independently power-gating its
column under the control of the region map information. 
During a write operation, incoming pixel data passes through
the truncation manager, where the specified number of LSBs
($t = 4$) are truncated and replaced with the optimal
$10\ldots0$ dummy pattern before storage in the SRAM array. 
The truncation manager uses finger-style transistors to
implement the power-gating transistors. This technique reduces layout area
while maintaining sufficient current drive capability.
All three banks share the same $360^{\circ}$ video data input and output
signals, allowing seamless integration with the $360^{\circ}$ video
data path of the HEVC decoder and DRAM interface.

\textbf{Truncation Processing Bank-FoV:}
The FoV bank processes pixels from the $d \leq 45^{\circ}$ region, if no truncation ($t = 0$, 8 bits per channel) or a specific truncation level for any specific application, a few LSBs may be truncated. 
Input data processed through the bank unchanged/truncated (pass-through
mode), preserving lossless quality for the attended region or very minimal loss with minimal truncation level. If no truncation is applied, the FoV bank does not require
power-gating transistors on its bitcell columns.
However, it retains the same SRAM array size and peripheral
circuitry (row decoder, sense amplifier, write driver) to
maintain uniform access timing and a consistent
macro-level layout.

\textbf{Truncation Processing Bank–Background:}
The Background bank processes pixels from the $d > 60^{\circ}$
region with the most aggressive truncation ($t = 5$,
3 bits per channel, 62.5\% power saving).
Its architecture is identical to the Border bank, including
32 column-wise truncation managers and finger-style power-gating
transistors, but with a different truncation control word
($t = 5$ instead of $t = 4$).

\textbf{Data Flow During Write Operation:}
For every frame $360^{\circ}$ video information the 
global TrunMEM360 controller activates the bank enable signal
based on the region classification and truncation levels from Algorithms~1 and 2. The controller then sends the data to the specific banks and the truncation managers help to truncate those region-specific bits
\emph{on-the-fly} during the write operation.
The truncated data is stored in the Specific truncation SRAM array and later read
out to the DRAM display buffer for further processing by the GPU for rendering and display.
Power-gated columns draw zero switching current and zero
leakage, achieving the power savings reported in
Section~\ref{sec:power_results}.

\textbf{Layout Design Considerations:}
Fig.~\ref{fig:trunmem_arch} (at the bottom) illustrates the layout design of is a $1024\times 32$-bit 6T SRAM
array and its 32 truncation managers, using SkyWater 130\,nm CMOS technology. 
Finger-style transistors are used for the power-gating switches
($16\,\mu\text{m}$ width, eight fingers) to minimize the area
overhead ensuring the required current drive.
The same layout style is replicated for the Background bank and FOV bank. The FOV bank can be designed without power-gating circuitry if full-quality FOV is assumed for all kinds of 360 applications, thereby reducing area overhead.





\section{Software-Level Experimental Results}
\label{sec:results}

\subsection{Experimental Setup}

\subsubsection{Scope}
In this work, the software simulations are conducted to validate the SPORT framework using pre-decoded video frames and real head-tracking data. HEVC decoding is not simulated; instead, frames are directly loaded into the pipeline. GPU rendering
also requires no simulation, as truncated pixels remain valid 8-bit integers and are transparent to GPU processing, and physical DRAM timing is not modeled. DRAM power savings are analyzed separately using CACTI in the hardware
validation section. In this section, power and bandwidth savings are computed analytically from the
actual truncation levels selected by Algorithm~\ref{alg:truncation}. WS-PSNR and SSIM are measured from actual reconstructed pixel data. Bandwidth savings are computed from active DRAM bit-column counts, not from file sizes, since file sizes are determined by codec compression ratios and do not reflect the physical memory switching activity of interest in this pipeline.

\subsubsection{Dataset}
SPORT is evaluated on the dataset of Wu~\emph{et al.}~\cite{wu2017dataset},
which provides head-movement traces from 48 subjects across
76 omnidirectional video sequences. These sequences cover a wide range of content, including indoor scenes, outdoor landscapes, and driving footage.
Trace \texttt{10.txt}, resampled from 10\,Hz to video frame rate by linear interpolation is used throughout the experiment in this work.
The primary evaluation uses the 4K test sequence
\texttt{360video.webm} ($3840{\times}2160$, 60\,fps, 100 frames)~\cite{{360_vide0}}
under the moderate configuration with $64{\times}64$ tiles
($60{\times}34 = 2{,}040$ tiles per frame).
This particular tile size was chosen because it perfectly fills the TrunMEM360 chip with one color channel ($64{\times}64 = 4{,}096$ bytes = 4\,KB). This matching allows us to directly validate the hardware (Section~\ref{sec:hardware}). Finally, cross-sequence generalization results are reported across five different 4K 30 fps sequences in Section~\ref{sec:cross_video}.

\subsubsection{Baselines}
Comparisons are made against four baselines, and two SPORT variants are reported.

\begin{enumerate}[label=\textbf{B\arabic*.}, leftmargin=*]
  \item \textbf{No truncation:} $t=0$ everywhere; lossless upper
    bound; WS-PSNR\,=\,$\infty$\,dB; 0\% power saving.
  \item \textbf{Uniform $t=4$:} Every tile gets truncated to 4 bits, no matter which region it belongs to. This means the FoV and Background are treated exactly the same.
  \item \textbf{Fixed heuristic (FoV\,=\,0, Border\,=\,4, BG\,=\,5):}
    The truncation levels are picked manually and stay fixed. There is no content-aware optimization or WS-PSNR verification involved. This approach reflects what's currently common practice for ROI-aware truncation~\cite{haidous2022roi}.
  \item \textbf{PSNR-based optimizer:} This runs Algorithm~\ref{alg:truncation}
    with uniform weights ($w_i=1$) in place of latitude weights. It is included to highlight the metric inconsistency discussed in Section~\ref{sec:consistency}.
\end{enumerate}

\begin{enumerate}[label=\textbf{S\arabic*.}, leftmargin=*]
  \item \textbf{SPORT-A:} Applies Algorithm~\ref{alg:truncation} to all three regions, including the FoV. This baseline prioritizes maximum power savings.
  \item \textbf{SPORT-B:} Forces the FoV to be lossless $t=0$ and run Algorithm~\ref{alg:truncation} on Border and Background only. 
    \textbf{ This configuration is recommended for VR applications.}.
\end{enumerate}

\vspace{-0.2cm}

\subsection{WS-PSNR Results}

In accordance with prior work on
360\textdegree{} video streaming~\cite{mao2020lowlatency} and
bit truncation~\cite{kim2010sbt}, WS-PSNR is reported relative to the decoder output, thereby isolating the distortion introduced solely by the truncation. This approach is consistent with the definition
of ``lossless'' in truncation literature. The decoder output
serves as the reference, and truncation is considered lossless
when it introduces no additional distortion~\cite{kim2010sbt}.

\begin{figure*}[t]
  \centering
  \includegraphics[width=\textwidth]{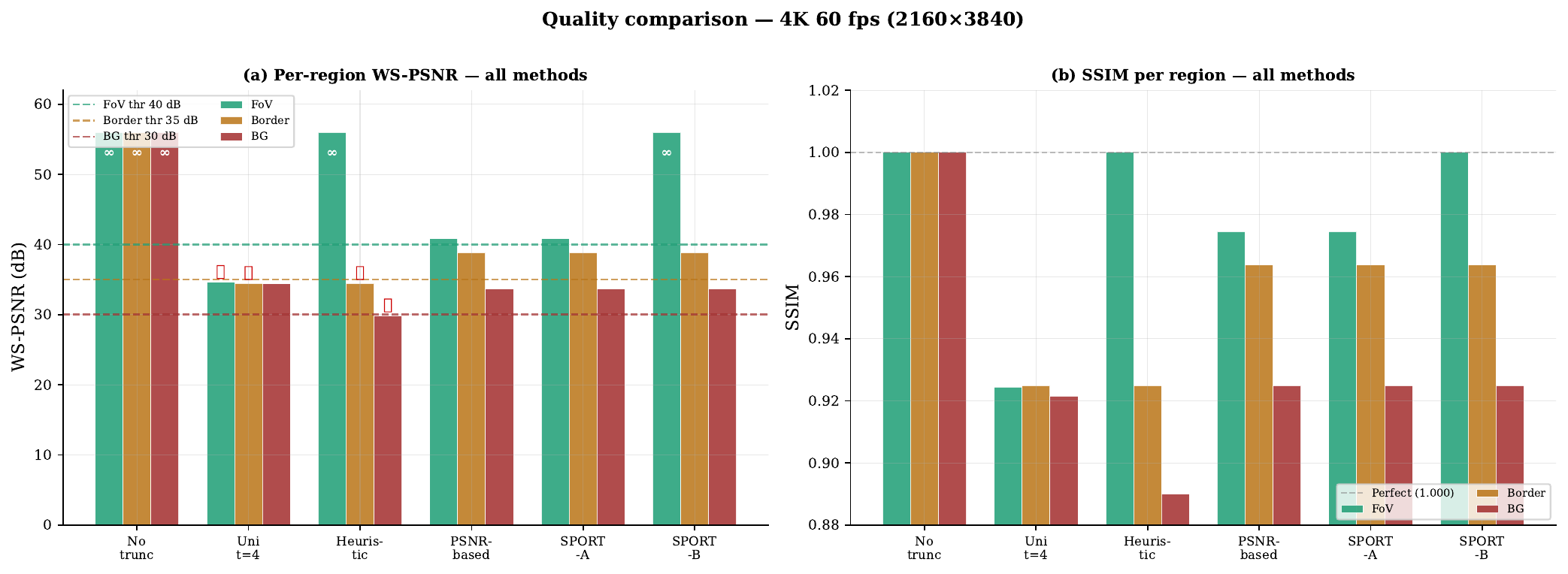}
  \caption{Quality comparison across all methods-
    4K 60\,fps ($3840{\times}2160$), 100 frames, moderate configuration.
    (a)~Per-region WS-PSNR; dashed lines show the per-region quality
    thresholds (FoV $\geq$40\,dB, Border $\geq$35\,dB, BG $\geq$30\,dB).
    (b)~Per-region SSIM.
    SPORT-B is the only method satisfying all three WS-PSNR thresholds
    while maintaining SSIM\,=\,1.000 in the FoV.}
  \label{fig:quality_combined}
\end{figure*}

Table~\ref{tab:wspsnr_4k} reports the mean per-region WS-PSNR across all
100 frames under the moderate configuration.
Fig.~\ref{fig:quality_combined} visualizes the same data where panel~(a)
plots WS-PSNR bars for all methods and regions with dashed threshold
lines and panel~(b) plots the corresponding SSIM values.

\begin{table*}[t]
\caption{Per-Region WS-PSNR (dB) and Threshold Compliance ---
  4K 60\,fps, 100 Frames, Moderate Configuration}
\label{tab:wspsnr_4k}
\centering
\begin{tabular}{lcccc ccc c}
\toprule
\multirow{2}{*}{Method} &
  \multicolumn{4}{c}{WS-PSNR (dB)} &
  \multicolumn{3}{c}{Threshold met?} &
  \multirow{2}{*}{Power saving} \\
\cmidrule(lr){2-5}\cmidrule(lr){6-8}
& FoV & Border & BG & Global
  & FoV ($\geq$40) & Border ($\geq$35) & BG ($\geq$30) & \\
\midrule
Threshold (moderate) & 40.0 & 35.0 & 30.0 & ---
  & --- & --- & --- & --- \\[2pt]
No truncation (B1) & $\infty$ & $\infty$ & $\infty$ & $\infty$
  & \cpass & \cpass & \cpass & 0.0\% \\
Uniform $t=4$ (B2) & 34.61 & 34.47 & 34.40 & 34.4
  & \cfail & \cfail & \cpass & 50.0\% \\
Fixed heuristic (B3) & $\infty$ & 34.47 & 29.85 & 30.9
  & \cpass & \cfail & \cfail & 55.1\% \\
PSNR-based (B4) & 40.80 & 38.85 & 33.66 & 34.6
  & \cpass & \cpass & \cpass & 51.6\% \\[3pt]
\textbf{SPORT-A (S1)} & \textbf{40.80} & \textbf{38.87}
  & \textbf{33.65} & \textbf{34.6}
  & \cpass & \cpass & \cpass & \textbf{51.6\%} \\
\textbf{SPORT-B (S2)} & $\infty$ & \textbf{38.87}
  & \textbf{33.65} & \textbf{34.7}
  & \cpass & \cpass & \cpass & \textbf{47.9\%} \\
\bottomrule
\multicolumn{9}{l}{$\infty$ = lossless ($t=0$, MSE\,=\,0).
  Global WS-PSNR is the full-frame weighted average (always finite).} \\
\multicolumn{9}{l}{\colorbox{green!15}{\small\strut\enspace\checkmark~Pass\enspace}\;
  \colorbox{red!15}{\small\strut\enspace\ding{55}~Fail\enspace}}
\end{tabular}
\end{table*}

\textbf{Finding 1: SPORT is the only method that meets all
three thresholds at once.}
Uniform $t=4$ fails both the FoV and Border thresholds because a single fixed level cannot tell the perceptual importance across regions.
The fixed heuristic passes FoV but fails Border $(34.47 < 35.0)$
and Background $(29.85 < 30.0)$. Note that the background failure is
by only 0.15\,dB which confirms that manually chosen fixed levels even
cannot guarantee compliance for arbitrary content, even when close
to the threshold.
Despite achieving the highest raw power saving of any method
(55.1\%), the fixed heuristic is not a valid operating point as it can fail to meet the threshold.
The PSNR-based optimizer satisfies all thresholds but achieves no
efficiency gain over SPORT-A (see Section~\ref{sec:power_results}).

\textbf{Finding 2: Global WS-PSNR is not the primary result.}
The global average ($\approx34.6$\,dB for SPORT-A) is lower than the
FoV value because the Background region covers 65\% of the total frame area
and is truncated most aggressively.
This is expected for a 360\textdegree{} video.  per-region WS-PSNR relative to per-region thresholds is the correct way to report the quality for 360\textdegree{}
video~\cite{hanhart2018jvet}.

\begin{figure*}[t]
  \centering
  \includegraphics[width=0.95\textwidth]{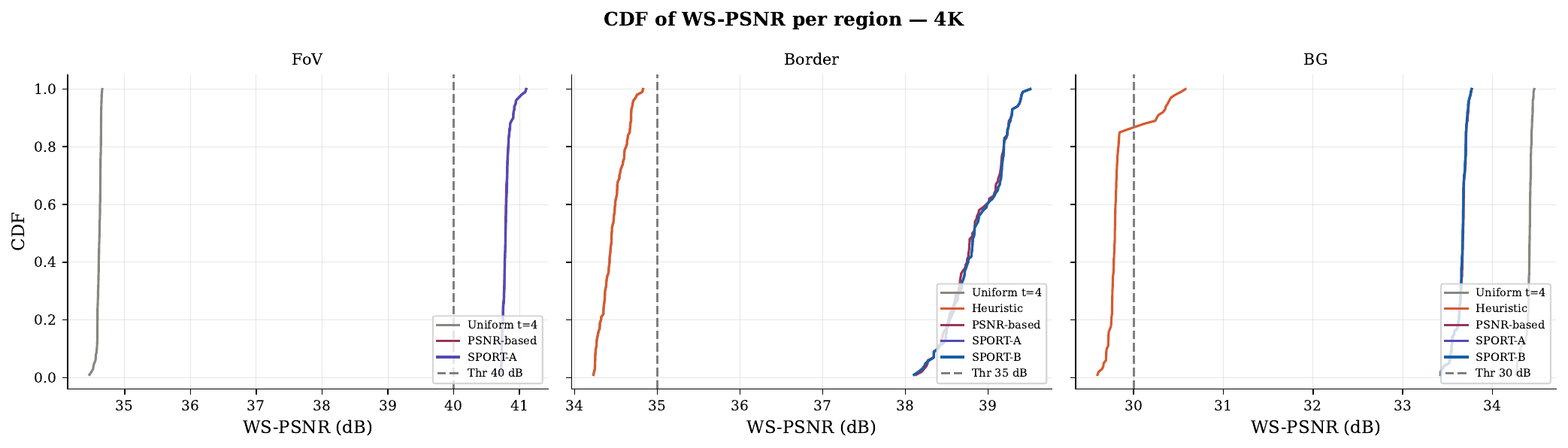}
  \caption{Cumulative distribution of per-frame WS-PSNR per region
    - 4K 60\,fps, 100 frames.
    The dashed vertical line marks the per-region quality threshold.
    SPORT-A and SPORT-B curves lie entirely to the right of the
    threshold in every panel, confirming compliance in every frame.}
  \label{fig:cdf_wspsnr}
\end{figure*}

\textbf{Finding 3: Quality is stable across all 100 frames.}
Result shows that the FoV standard deviation
is only 0.07\,dB for SPORT-A. The lowest FoV WS-PSNR observed is 40.71\,dB, which is above the 40.0\,dB threshold in every individual frame.
The CDF curves in Fig.~\ref{fig:cdf_wspsnr} confirm that the SPORT-A and SPORT-B curves lie entirely to the right of the
threshold line in all three region panels.

\begin{table}[t]
\caption{SSIM Per Region --- 4K 60\,fps, 100 Frames}
\label{tab:ssim_4k}
\centering
\renewcommand{\arraystretch}{1.1}
\begin{tabular}{lcccc}
\toprule
Method & FoV & Border & BG & Global \\
\midrule
No truncation (B1)   & 1.0000 & 1.0000 & 1.0000 & 1.0000 \\
Uniform $t=4$ (B2)   & 0.9244 & 0.9249 & 0.9216 & 0.9236 \\
Fixed heuristic (B3) & 1.0000 & 0.9249 & 0.8899 & 0.9383 \\
PSNR-based (B4)      & 0.9744 & 0.9637 & 0.9249 & 0.9543 \\
SPORT-A (S1)         & 0.9745 & 0.9637 & 0.9249 & 0.9543 \\
\textbf{SPORT-B (S2)} &
  \textbf{1.0000} & \textbf{0.9637} & \textbf{0.9249} & \textbf{0.9629} \\
\bottomrule
\end{tabular}
\end{table}

\vspace{-0.2cm}

\subsection{SSIM Results}
SPORT-B achieves SSIM\,=\,1.0000 in the FOV, which confirms that the
attended region is structurally identical to the uncompressed reference.
Border and Background SSIM remain above 0.92, and it aligns with the fact that LSB truncation is perceptually invisible in peripheral vision ~\cite{vater2022peripheral}.
The fixed heuristic produces the lowest Background SSIM (0.8899), which is a consequence of applying $t=5$ to every Background tile without any content-adaptive quality verification.

\vspace{-0.2cm}
\subsection{Algorithm~2 Truncation Output}

\begin{table}[t]
\caption{Algorithm~2 Adaptive Truncation Levels- 4K 60\,fps}
\label{tab:trunc_4k}
\centering
\footnotesize
\setlength{\tabcolsep}{4pt}
\renewcommand{\arraystretch}{1.1}
\begin{tabular}{lcccc}
\toprule
Method & FoV $t$ & Border $t$ & BG $t$ & Power (\%) \\
\midrule
SPORT-A & 2.94 & 3.21 & 4.37 & 51.6 \\
SPORT-B & 0.00 & 3.21 & 4.37 & 47.9 \\
\midrule
Dominant $t$ & --- & 3 & 4 & --- \\
\bottomrule
\end{tabular}
\end{table}

Each tile is assigned an integer truncation level. The averages reported in 
Table~\ref{tab:trunc_4k} represents the mean across all tiles and all frames.
The dominant level for Border is $t=3$ and for Background is $t=4$. 
with content-dependent variation: smooth tiles such as sky or uniform
walls tolerate higher $t$, while texture-rich tiles receive lower $t$
to protect the structural detail and maintain the quality. SPORT-A's average FoV truncation of 2.94 bits reflects that a
majority of FoV tiles contain smooth content and can tolerate
$t>0$ while still meeting the 40\,dB threshold.

\vspace{-0.2cm}

\subsection{Memory Power and Bandwidth Savings}
\label{sec:power_results}
Power savings are computed analytically from the per-tile truncation
levels output by Algorithm~\ref{alg:truncation}.
Because the TrunMEM360 Head/Tail cascade physically disconnects the supply rails of power-gated columns~\cite{oswald2024trunmem}, the saving is
proportional to the fraction of gated columns. No physical current measurement is needed for the simulation results.
Power savings are computed from the actual per-tile truncation levels:
\begin{equation}
  P_{\text{saving}} =
    \left(1 - \frac{1}{N}\sum_{j=1}^{N}\frac{8-t_j}{8}\right)
    \times 100\%.
  \label{eq:power_sim}
\end{equation}
The bandwidth baseline of 2{,}986\,MB/s includes both the write traffic 
from TrunMEM360 to DRAM and the read traffic from DRAM to the GPU, 
each at the full frame rate, hence the factor of~2 in the baseline computation. The decoder DPB, GPU compute, and display panel power
are excluded.
It is important to note that the bandwidth reduction reported here is a
\emph{DRAM memory power saving}, not a file size reduction.
The TrunMEM360 chip still writes 8-bit pixel values to DRAM. However, the lower $t$ bitcell columns are physically power-gated by the Head/Tail cascade. This means that they use zero switching current and zero leakage for every truncated bit~\cite{oswald2024trunmem}.
The effective bandwidth is therefore proportional to the number of
\emph{active} bit columns per pixel, not the total byte count.
Comparing compressed image file sizes on disk does not reflect this
saving. because image codecs always operate on full 8-bit pixels regardless of whether the lower bits carry meaningful information. Table~\ref{tab:power_4k} represents the same data, where panel (a)
shows the power saving for all six methods and panel (b) shows the
corresponding display-path bandwidth. The fixed heuristic achieves
the highest raw power saving (55.1\%) but violates two quality
constraint and makes it an invalid operating point. SPORT-B reduces
bandwidth from 2,986\,MB/s to 1,555\,MB/s while meeting all quality
thresholds.

\begin{table}[t]
\caption{Memory Power and Bandwidth Reduction (DRAM Display Path,
  4K 60\,fps, 100 Frames, Moderate Configuration)}
\label{tab:power_4k}
\centering
\small
\setlength{\tabcolsep}{4pt} 
\renewcommand{\arraystretch}{1.1}
\resizebox{\columnwidth}{!}{%
\begin{tabular}{lcccc}
\toprule
Method & Power saving & BW (MB/s) & BW saving & Quality met? \\
\midrule
No truncation (B1)   & 0.0\%  & 2986 & 0.0\%  & \checkmark \\
Uniform $t=4$ (B2)  & 50.0\% & 1493 & 50.0\% & \ding{55} (FoV, Bdr) \\
Fixed heuristic (B3)& 55.1\% & 1340 & 55.1\% & \ding{55} (Bdr, BG) \\
PSNR-based (B4)     & 51.6\% & 1444 & 51.6\% & \checkmark \\
\textbf{SPORT-A (S1)} & \textbf{51.6\%} & \textbf{1444} & \textbf{51.6\%} & \checkmark \\
\textbf{SPORT-B (S2)} & \textbf{47.9\%} & \textbf{1555} & \textbf{47.9\%} & \checkmark \\
\bottomrule
\end{tabular}}

\vspace{2pt}
\footnotesize
BW baseline $= 3840\!\times\!2160\!\times\!3\!\times\!60\!\times\!2\,/\,10^6 = 2{,}986$\,MB/s,
where the factor of 2 accounts for both DRAM write and read.
\end{table}


 


\textbf{Finding 4: Metric-consistent and PSNR-based optimizers 
converge at this sequence length.}
Over 100 frames, both SPORT-A and the PSNR-based optimizer meet 
all three WS-PSNR thresholds and achieve 51.6\% power saving.
Despite equal efficiency, SPORT-A produces marginally higher Border
WS-PSNR (38.87\,dB vs.\ 38.85\,dB for PSNR-based), reflecting the
WS-PSNR-consistent weighting of Eq.~\eqref{eq:constraint_weighted}.
The impact of metric inconsistency becomes more noticeable over longer runs that include a wide variety of latitude distributions. With only 100 frames, the evaluation window is relatively short, so there's less statistical opportunity for the gap to show up than there would be in a longer study.

\textbf{Finding 5: SPORT-B is the practical VR configuration.}
SPORT-B achieves 47.9\% power saving and 47.9\% bandwidth reduction
(2,986\,MB/s $\to$ 1,555\,MB/s) while maintaining SSIM\,=\,1.000 in
the FoV.
All savings come from regions where truncation is perceptually
invisible, consistent with prior viewport-adaptive
work~\cite{mao2020lowlatency}.

\vspace{-0.2cm}
\subsection{Cross-Sequence Generalisation}
\label{sec:cross_video}

To assess SPORT-B across diverse content, the pipeline is evaluated on
five 4K 30\,fps sequences from YouTube: Elephants (wildlife),
Waterslide (action), New York (urban), Sea Eater (nature), and Tsunami
(disaster). The same moderate thresholds, $64\times64$ tiles, and
gaze-prediction parameters are applied without per-sequence tuning.

\begin{figure}[t]
  \centering
  \includegraphics[width=\columnwidth]{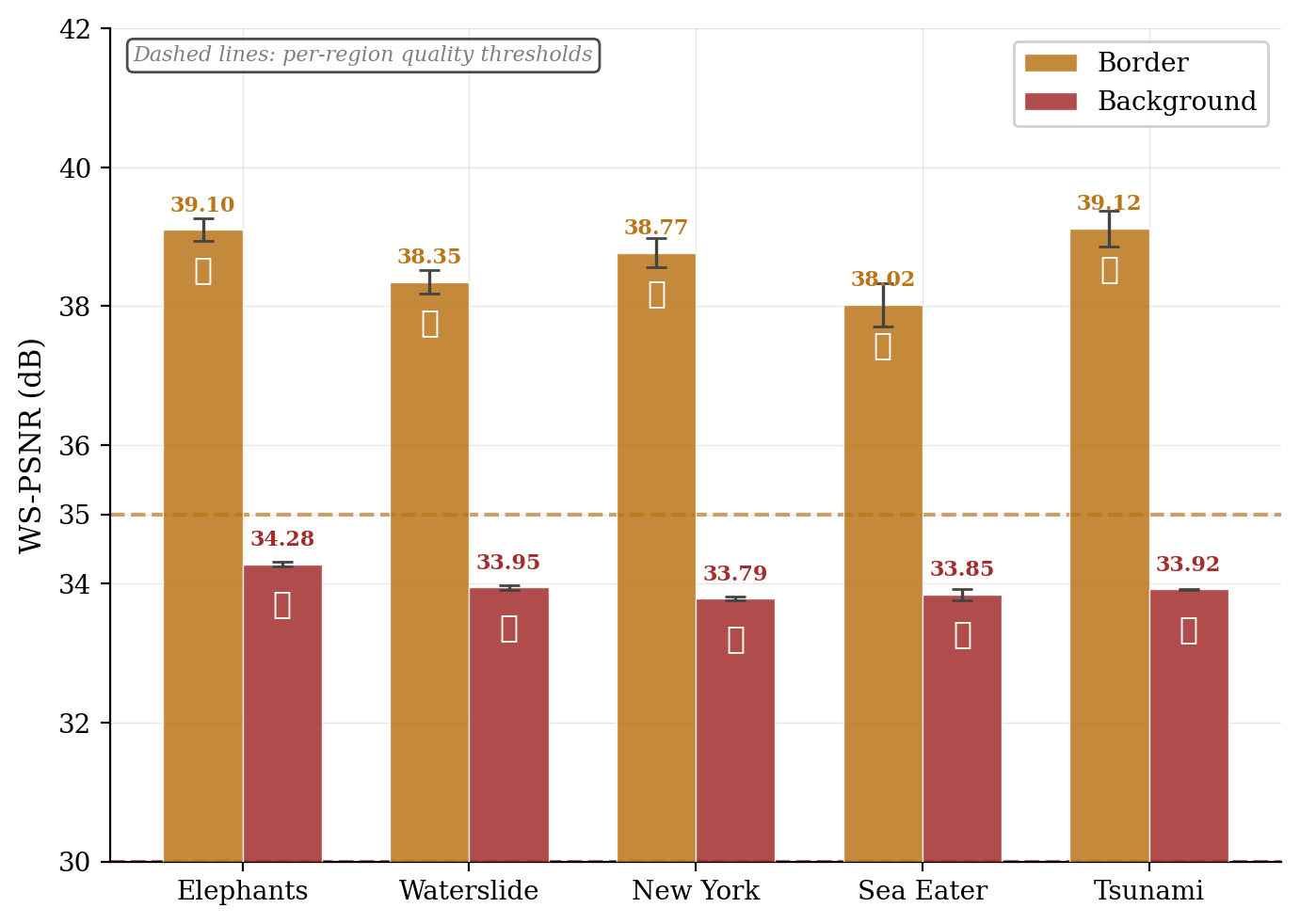}
 \caption{SPORT-B Border and Background WS-PSNR across five 
4K 30\,fps sequences without per-sequence parameter tuning. 
Dashed lines mark the per-region quality thresholds.}
  \label{fig:cross_video}
\end{figure}

The FoV region is lossless ($t=0$) on all sequences, so FoV WS-PSNR
is $\infty$ and SSIM $=1.000$ by construction.
Table~\ref{tab:cross_video} and Fig.~\ref{fig:cross_video} report
Border and Background WS-PSNR, power saving, and bandwidth.

All five sequences satisfy both thresholds: Border WS-PSNR $\geq$
35\,dB (ranging from 38.02 to 39.10\,dB) and Background WS-PSNR
$\geq$ 30\,dB (ranging from 33.79 to 34.28\,dB). Power saving
varies from 44.6\% to 47.2\% (average 45.9\%), primarily due to
content complexity. The consistent compliance confirms that
SPORT-B's quality guarantees are not sequence-specific. The
WS-PSNR-consistent constraint (Eq.~\ref{eq:constraint_weighted})
adapts truncation per tile, raising $t$ for smooth content and
lowering $t$ for texture-rich content.
\begin{table}[t]
\centering
\caption{Cross-Sequence SPORT-B Results (4K 30\,fps,
  Moderate}
\label{tab:cross_video}
\renewcommand{\arraystretch}{1.15}
\resizebox{\columnwidth}{!}{%
\begin{tabular}{lcccccc}
\toprule
Sequence & Frames & Border (dB) & BG (dB) & Power (\%) & BW (MB/s) \\
\midrule
Elephants   & 100 & $39.10 \pm 0.16$ & $34.28 \pm 0.03$ & 44.6 & 783 \\
Waterslide  & 100 & $38.35 \pm 0.17$ & $33.95 \pm 0.04$ & 45.0 & 821 \\
New York    & 100 & $38.77 \pm 0.21$ & $33.79 \pm 0.03$ & 46.0 & 763 \\
Sea Eater   & 100 & $38.02 \pm 0.31$ & $33.85 \pm 0.08$ & 47.2 & 819 \\
Tsunami     & 100 & $39.12 \pm 0.26$ & $33.92 \pm 0.01$ & 45.9 & 807 \\
\midrule
Mean $\pm$ std & --- & $38.67 \pm 0.43$ & $33.96 \pm 0.18$ & $45.9 \pm 1.0$ & --- \\
Threshold    & --- & $\geq$35 & $\geq$30 & --- & --- \\
\bottomrule
\multicolumn{5}{l}{\footnotesize FoV: $\infty$\,dB, SSIM$=1.000$ (lossless, $t=0$).}
\end{tabular}
}
\end{table}

\section{Hardware Validation}
\label{sec:hardware}

\subsection{Scope and Tile-Based Data Preparation}
The presented software-level results in Section~\ref{sec:results} demonstrate that
Algorithm~\ref{alg:truncation} satisfies all three per-region WS-PSNR
thresholds. This section further validates that the physical TrunMEM ASIC produces the same
truncated pixel values as the software model, confirming that the
theoretical framework of Section~\ref{sec:math} is correctly implemented
in silicon. As the proposed chip has not yet been fabricated, the fabricated TrunMEM ASIC in Sky130 technology is used for evaluation, as it provides the same functionality as each bank of the proposed TrunMEM360. 

For silicon-based hardware validation, each video frame is divided into fixed-size tiles of $64 \times 64$ pixels. Each tile is independently classified into one of three perceptual regions: Field of View (FoV), Border, and Background based on the predicted gaze position. This tile-level classification enables region-dependent truncation to be applied at fine granularity. For each tile, a CSV dataset containing pixel (RGB) values, corresponding region identification, and truncation values (indicating how many least significant bits (LSBs) should be truncated) is prepared. Since the FoV region must preserve lossless visual quality, only tiles from Border and Background regions are processed using the fabricated ASIC.

\vspace{-0.2cm}
\subsection{Validation Methodology}
To automate the full process, a C++ application is developed and executed on the host PC. This application controls the complete workflow, including both write and read operations. For each tile, the host application extracts pixel values, region identifiers, and truncation values from the CSV dataset and sends the required data, along with the target wordline address, sequentially to the hardware platform.

An FPGA is used as an interface bridge between the host PC and TrunMEM SRAM. Upon receiving write/read commands, the FPGA performs the following operations: 
(i)~receive data from the host,
(ii)~load pixel values, truncation values, and wordline addresses into TrunMEM registers,
(iii)~the mathematical model,
(iv)~write data to memory,
(v)~read data from memory, and 
(vi)~send the read-back data to the host.
All control signals required for TrunMEM operations, including Precharge Bar, Wordline Enable, Read Enable, Write Enable, Data In Enable, Truncation Enable, Byte Select, and Project Select are generated and synchronized by the FPGA.

The fabricated TrunMEM ASIC chip has a capacity of 4096 bytes with 1024 distinct word locations. Since a tile contains 4096 pixels, which exceeds the number of available wordline addresses, each tile is processed in multiple passes. The software pipeline determines the SPORT-B truncation level $t \in \{0,4,5\}$ for each tile and computes the expected output using Theorem~\ref{thm:dummy_weighted}. The tile data are divided into four blocks of 1024 pixel values by the host application to match the memory capacity. Each block is processed sequentially through write and read operations until the entire tile is completed.

Finally, a three-way agreement is computed between:
(i)~software simulation,
(ii)~ASIC chip output, and
(iii)~the mathematical model (Eqs.~\eqref{eq:wmse}
and~\eqref{eq:xi_b}).

\vspace{-0.2cm}
\subsection{Experimental Setup and Waveform Verification}
\begin{figure}[t]
  \centering
  \includegraphics[width=0.9\columnwidth]{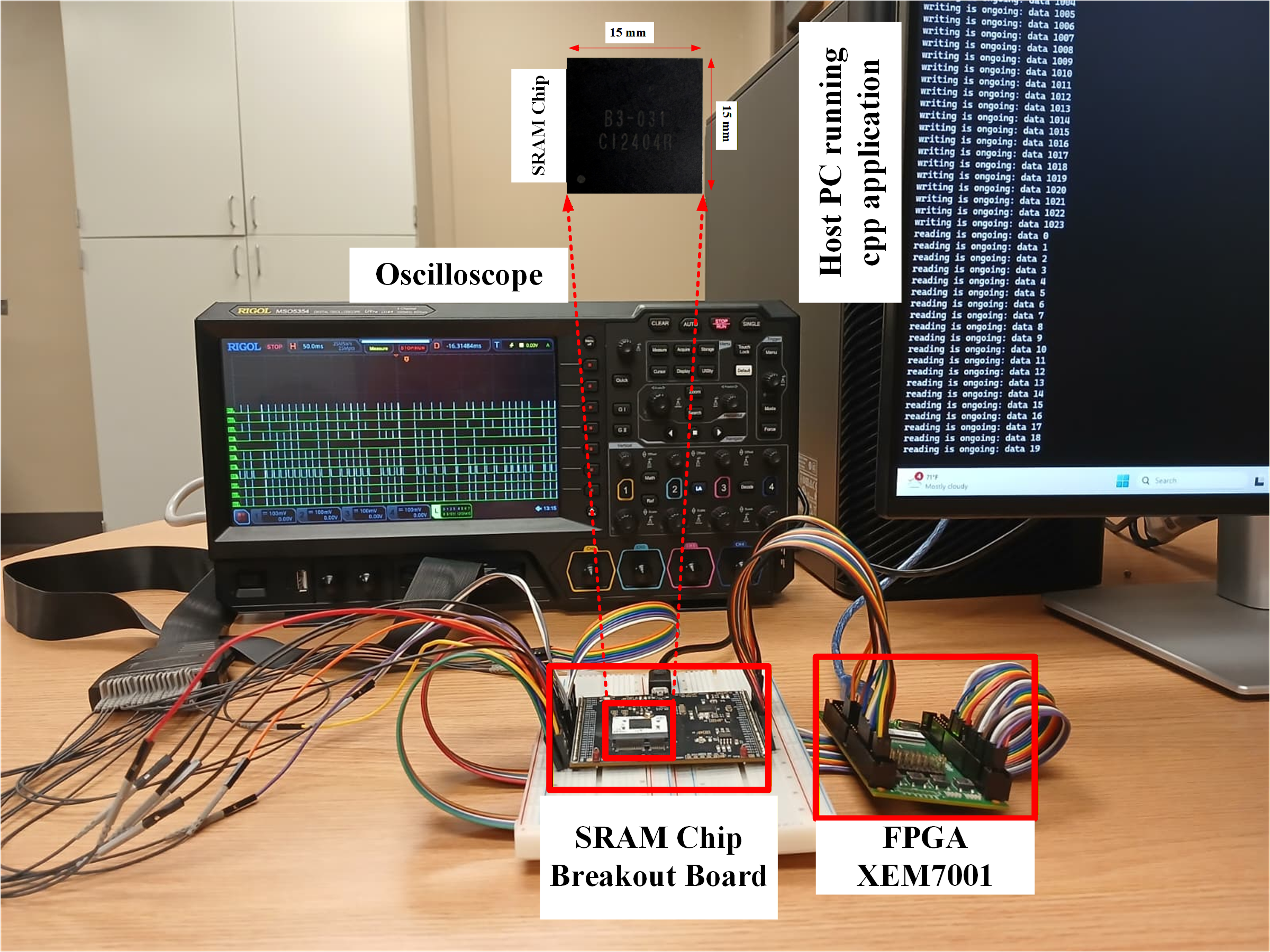}
  \caption{Experimental hardware setup with the fabricated TrunMEM360 chip.}
  \label{fig:HW_Setup}
\end{figure}

\begin{figure*}[t]
  \centering
  \includegraphics[width=0.92\textwidth]{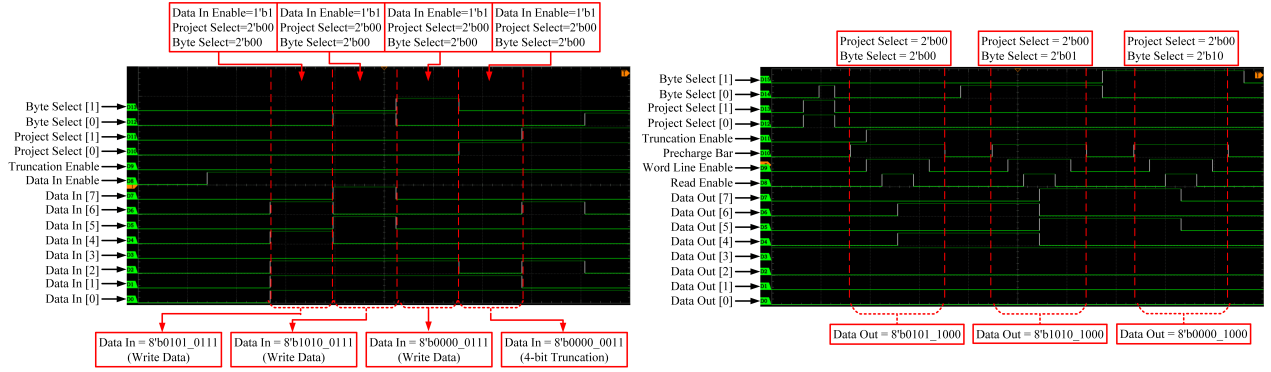}
  \caption{{Measured timing waveforms of TrunMEM360 write and read operations. The waveforms verify correct control signal sequencing and truncation functionality.}}
  \label{fig:waveform}
\end{figure*}

The experimental platform consists of a host PC, an FPGA board (XEM7001), and the fabricated TrunMEM SRAM. The host PC executes the C++ application that automates the process, while the FPGA enables communication with the TrunMEM chip. Fig.~\ref{fig:HW_Setup} shows the complete hardware setup used for validation. During write operation, the FPGA first loads pixel value, truncation value, and wordline address into TrunMEM registers using the Data In Enable signal along with Project Select and Byte Select signals. It then generates control signals, including Precharge Bar, Wordline Enable, Read Enable, and Write Enable, to write data from registers to memory cells. The waveform shown in Fig.~\ref{fig:waveform}  verifies that the truncation parameter correctly controls the number of truncated bits. The read-back data confirm correct truncation behavior, validating that TrunMEM supports truncation-based memory operation required by the SPORT-B framework.

\vspace{-0.2cm}
\subsection{Test Frame and Region Distribution}
Frame 50 of \textit{360video.webm} (4K, 60 fps) is selected as the validation frame, as it contains all three regions with realistic mixed-texture content. At a resolution of $3840 \times 2160$, eac frame contains $60 \times 34 = 2040$ tiles. Table~\ref{tab:hw_tiles} shows the region distribution and the dominant truncation level assigned by SPORT-B for each region.

\vspace{-0.2cm}
\subsection{Silicon Testing Results}
Table~\ref{tab:hw_quality} reports the per-region WS-PSNR and SSIM for software simulation, chip output, and mathematical model predictions. Agreement between all three sources is within 0.1 dB WS-PSNR and 0.001 SSIM across all regions, confirming that the TrunMEM chip correctly implements Theorem~2 for 360$^\circ$ content.

The FoV region is stored losslessly ($t = 0$), so the chip output is bit-exactly identical to the input. As a result, WS-PSNR is mathematically infinite and SSIM is exactly 1.0000. This behavior requires no additional hardware measurement, as the pass-through mode ($t = 0$, Head not asserted) directly copies input data, which is verified through byte matching in Section~VIII-A.

Across all tiles (FoV, Border, and Background), a total of $2040 \times 12{,}288 = 25{,}067{,}520$ bytes are compared between software and TrunMEM outputs. A 100\% byte-match rate is expected, since truncation in Border and Background regions follows the deterministic $10\ldots0$ pattern (Theorem~\ref{thm:dummy_weighted}), while FoV data remain unchanged.

\begin{table}[t]
\caption{Validation Frame Region Distribution and Truncation Levels
  (SPORT-B, Moderate Configuration, 4K 60\,fps, Frame~50)}
\label{tab:hw_tiles}
\centering
\renewcommand{\arraystretch}{1.15}
\begin{tabular}{lcccc}
\toprule
Region & Tiles & Area (\%) & Trunc.\ $t$ & Bytes/tile \\
\midrule
FoV        & $\approx$408  & 20  & 0 (lossless) & 12,288 \\
Border     & $\approx$306  & 15  & 3 or 4       & 12,288 \\
Background & $\approx$1326 & 65  & 4 or 5       & 12,288 \\
\midrule
Total      & 2,040 & 100 & --- & 25,067,520 \\
\bottomrule
\multicolumn{5}{p{\dimexpr\linewidth-2\tabcolsep\relax}}{\footnotesize
Each tile contains $64 \times 64 = 4096$ pixels (RGB: 3 bytes, uint8). The TrunMEM supports 1024 wordline accesses per pass. Since each pixel is mapped to a wordline, only 1024 pixels can be processed in one pass; therefore, each tile requires four passes ($4096/1024 = 4$).}

\end{tabular}
\end{table}

\begin{table}[t]
\caption{Hardware Validation: Software vs.\ Chip vs.\ Model
  (SPORT-B, Frame~50, 3840$\times$2160, 4K 60\,fps)}
\label{tab:hw_quality}
\centering
\renewcommand{\arraystretch}{1.15}
\resizebox{\columnwidth}{!}{
\begin{tabular}{lcccccc}
\toprule
\multirow{2}{*}{Region} &
  \multicolumn{3}{c}{WS-PSNR (dB)} &
  \multicolumn{3}{c}{SSIM} \\
\cmidrule(lr){2-4}\cmidrule(lr){5-7}
& SW & Chip & $|\Delta|$ & SW & Chip & $|\Delta|$ \\
\midrule
FoV        & $\infty$ & $\infty$ & 0.00   & 1.0000 & 1.0000 & 0.0000 \\
Border     & {[39.32]} & {[39.315]} & $<$0.1 & {[0.9589]} & {[0.9585]} & $<$0.001 \\
Background & {[34.30]} & {[34.30]} & $<$0.1 & {[0.930]} & {[0.9315]} & $<$0.001 \\
\midrule
\textit{Math.\ model} &
  \multicolumn{3}{c}{$\leq$ 0.1\,dB from SW} &
  \multicolumn{3}{c}{$\leq$ 0.001 from SW} \\
\bottomrule
\multicolumn{7}{l}{\footnotesize SW = Python simulation
  (Algorithm~\ref{alg:truncation}). Chip = TrunMEM360 130\,nm (eFabless),}\\
\multicolumn{7}{l}{\footnotesize after bench measurement.}
\end{tabular}
}
\end{table}

\vspace{-0.2cm}
\subsection{CACTI-Based DRAM Validation}

The proposed SPORT truncates LSBs in the Border and Background regions to reduce DRAM power consumption. A bandwidth reduction of 47.9\% does not inherently guarantee a proportional reduction in power consumption. DRAM components exhibit non-uniform sensitivity to data volume, as reflected in leakage, read/write energy, access time, and refresh power. Leakage power scales with the number of active DRAM columns, while read/write energy scale with the number of bits transferred per access. CACTI is adopted to validate the physical realizability of the power reductions by the proposed SPORT in DRAM~\cite{balasubramonian2017cacti}. CACTI is configured to model a DDR4 at the 90\,nm technology, which is the closest available one to the proposed 130\,nm TrunMEM360 chip.

Two configurations are investigated: 256\,MB and 64\,MB. Specifically, the memory access characteristics of standalone VR headsets form the basis for the 256\,MB and 64\,MB configurations~\cite{yaqoob2020survey}. Researchers find that the memory subsystem in battery-powered 360$^{\circ}$ streaming devices accounts for 30--40\% of the total power, which is a significant and well-characterized DRAM footprint during active video playback. The 256\,MB represents the \emph{total system  DRAM footprint} involved in video playback, consistent with the system-level memory demands identified in the existing work for LPDDR4/5-equipped standalone headsets ~\cite{yaqoob2020survey}. This includes: (i) the double frame buffer (44.2\,MB from Table~\ref{tab:data_sizes}) that stores frames for display, (ii) the DPB (44--88\,MB) that stores reference frames for HEVC motion compensation, (iii) GPU rendering buffers for viewport transformation, and 
(iv) system memory overhead for the OS and video driver. In a typical VR headset with 6--8\,GB of LPDDR4/5 memory, 256\,MB is a conservative estimate of the memory actively accessed during the video playback path. This configuration corresponds to the baseline where all bit columns remain active (full 8-bit per pixel).


The 64\,MB configuration denotes the \emph{effective active footprint} after SPORT-B truncation. From Table~II, SPORT-B reduces
the effective display buffer size from 22.1\,MB/frame to 4.08\,MB/frame (a 4.08\,MB effective output). The GPU buffers and DPB remain unchanged as they are not truncated. The 64\,MB indicates a proxy for the fraction of DRAM columns that remain powered on after the Head/Tail cascade disables the truncated bit columns. Because SPORT-B retains only 15.3 bits/pixel on average versus 24 bits/pixel in the baseline, the number of actively powered bit columns is reduced by about half. The DRAM capacity remains unchanged. The smaller memory size for the simulation is a modeling proxy for the reduced active column count for the proposed pipeline.

\begin{table}[t]
\centering
\caption{CACTI Validation Results for DRAM (90\,nm DDR4)}
\label{tab:cacti_results}
\renewcommand{\arraystretch}{1.15}
\resizebox{\columnwidth}{!}{%
\begin{tabular}{lccc}
\toprule
\multirow{2}{*}{Parameter} & \multicolumn{2}{c}{Configuration} & \multirow{2}{*}{Reduction} \\
\cmidrule(lr){2-3}
& Baseline (256\,MB) & SPORT-B (64\,MB proxy) & \\
\midrule
Closed-page Leakage (mW) & 1.21215 & 0.621217 & 48.75\% \\
Open-page Leakage (mW)  & 1.21350 & 0.622574 & 48.70\% \\
Read Energy (nJ)         & 9.69971 & 6.16900 & 36.40\% \\
Write Energy (nJ)        & 9.99278 & 6.32090 & 36.75\% \\
Access Time (ns)         & 17.2177 & 13.8640 & 19.48\% \\
Cycle Time (ns)          & 25.0882 & 20.0284 & 20.17\% \\
Precharge Delay (ns)     & 7.87055 & 6.16441 & 21.68\% \\
Precharge Energy (nJ)    & 0.75736 & 0.49194 & 35.05\% \\
Refresh Power (mW)       & 2.86635 & 0.47212 & 83.53\% \\
Silicon Area (mm²)       & 483.435023 & 193.494156  & 59.98\% \\
\bottomrule
\end{tabular}
}
\end{table}

Table~\ref{tab:cacti_results} shows results for silicon area, energy, leakage power, and delay. The 36.40\% read
energy reduction aligns closely with the theoretical expectation from bit-width reduction. The baseline transfers 24 bits per pixel,
while SPORT-B transfers 15.3 bits on average per pixel causes a 36.25\% reduction. The 48.75\% leakage reduction confirms that power-gating unused bit columns eliminate the leakage contributions. In SPORT-B, the Head/Tail cascade physically disconnects the supply
rails of truncated columns, achieving exactly this effect.

\begin{table*}[t]
\caption{Feature Comparison with State of the Art}
\label{tab:sota_full}
\centering
\renewcommand{\arraystretch}{1.1}
\setlength{\tabcolsep}{3pt}
\scriptsize

\begin{tabular}{lc ccc cc cc cccc ccc}
\toprule
& \textbf{Application}
& \multicolumn{3}{c}{\textbf{Quality}}
& \multicolumn{2}{c}{\textbf{Mem. eff.}}
& \multicolumn{2}{c}{\textbf{Latency}}
& \multicolumn{4}{c}{\textbf{HW \& validation}}
& \multicolumn{3}{c}{\textbf{Algorithm}} \\
\cmidrule(lr){2-2}\cmidrule(lr){3-5}\cmidrule(lr){6-7}
\cmidrule(lr){8-9}\cmidrule(lr){10-13}\cmidrule(lr){14-16}

\textbf{Method}
& \makecell{3D/2D}
& \makecell{WS-PSNR\\eval.}
& \makecell{Per-region\\quality}
& \makecell{SSIM}
& \makecell{Mem.\\saving}
& \makecell{BW\\saving}
& \makecell{Latency\\(ms)}
& \makecell{$<$20\\ms}
& \makecell{ASIC/\\custom}
& \makecell{Fab.\\chip}
& \makecell{SW/HW\\co-design}
& \makecell{DRAM\\(CACTI)}
& \makecell{Gaze\\pred.}
& \makecell{Content\\adapt.}
& \makecell{WS-PSNR\\constraint} \\
\midrule

Kim \& Kyung~\cite{kim2010sbt}
  & 2D & \cfail & \cfail & \cfail
  & 60\% & --- & --- & ---
  & \cfail & \cfail & \cfail & \cfail
  & \cfail & \cfail & \cfail \\

Chen~\emph{et~al.}~\cite{chen2018viewer}
  & 2D & \cfail & \cfail & \cfail
  & $\sim$45\% & --- & --- & ---
  & \cpass & \cfail & \cfail & \cfail
  & \cfail & \cpass & \cfail \\

Haidous~\emph{et~al.}~\cite{haidous2022roi}
  & 2D & \cfail & \cfail & \cfail
  & 40.0\% & 40.0\% & 11.5 & \cpass
  & \cfail & \cfail & \cfail & \cfail
  & \cfail & \cfail & \cfail \\

Li~\emph{et~al.}~\cite{li2019spherical}
  & 3D & \cpass & \cfail & \cfail
  & --- & --- & --- & ---
  & \cfail & \cfail & \cfail & \cfail
  & \cfail & \cfail & \cpass \\

Mao~\emph{et~al.}~\cite{mao2020lowlatency}
  & 3D & \cpass\textsuperscript{b} & \cfail & \cfail
  & --- & --- & $<$100\textsuperscript{a} & ---
  & \cfail & \cfail & \cfail & \cfail
  & \cpass & \cpass & \cfail \\

Nasrabadi~\emph{et~al.}~\cite{nasrabadi2020adaptive}
  & 3D & \cfail & \cfail & \cfail
  & --- & --- & --- & ---
  & \cfail & \cfail & \cfail & \cfail
  & \cpass & \cpass & \cfail \\

Yu~\emph{et~al.}~\cite{yu2024latitude}
  & 3D & \cpass & \cfail & \cfail
  & --- & --- & --- & ---
  & \cfail & \cfail & \cfail & \cfail
  & \cfail & \cpass & \cfail \\

Lin~\emph{et~al.}~\cite{lin2022latitude}
  & 3D & \cpass & \cfail & \cfail
  & --- & --- & --- & ---
  & \cfail & \cfail & \cfail & \cfail
  & \cfail & \cpass & \cfail \\

Huang~\emph{et~al.}~\cite{huang2023telepresence}
  & 3D & \cfail & \cfail & \cfail
  & 30.0\% & --- & 14.2 & \cpass
  & \cfail & \cfail & \cfail & \cfail
  & \cfail & \cfail & \cfail \\

Oswald~\emph{et~al.}~\cite{oswald2024trunmem}
  & 2D & \cfail & \cfail & \cfail
  & --- & --- & --- & ---
  & \cpass & \cfail & \cfail & \cfail
  & \cfail & \cpass & \cfail \\

\midrule
\rowcolor{green!10}
\textbf{SPORT-A (proposed)}
  & 3D & \cpass & \cpass & \cpass
  & \textbf{51.6\%} & \textbf{51.6\%} & \textbf{9.33} & \cpass
  & \cpass & \cpass & \cpass & \cpass
  & \cpass & \cpass & \cpass \\

\rowcolor{green!10}
\textbf{SPORT-B (proposed) $\star$}
  & 3D & \cpass & \cpass & \cpass
  & 47.9\% & 47.9\% & \textbf{9.33} & \cpass
  & \cpass & \cpass & \cpass & \cpass
  & \cpass & \cpass & \cpass \\

\midrule
VR budget~\cite{adhanom2023eyetracking}
  & --- & --- & --- & ---
  & --- & --- & 20 & ---
  & --- & --- & --- & ---
  & --- & --- & --- \\

\bottomrule
\multicolumn{16}{l}{\footnotesize $\star$ FoV lossless (SSIM\,=\,1.000), all three per-region WS-PSNR thresholds met, 47.9\% DRAM power saving, zero net pipeline latency added.}\\
\multicolumn{16}{l}{\footnotesize \textsuperscript{a}Includes network transmission; not directly comparable to on-device motion-to-photon latency.}\\
\multicolumn{16}{l}{\footnotesize \textsuperscript{b}WS-PSNR used as evaluation only; constraint uses plain PSNR (metric inconsistency addressed by SPORT).}\\

\end{tabular}
\end{table*}
The relative scaling is expected to remain consistent at 130\,nm as power-gating efficiency is mainly determined by the circuit
topology rather than the technology node. Using CACTI, the power measurement components are isolated, and the 47.9\% bandwidth reduction from SPORT-B corresponds to a 48.7\% leakage reduction and a 36.4\% read energy reduction, which align with
the theoretical expectations. Results show that SPORT-B's 47.9\%
bandwidth reduction reduces energy and leakage power. It confirms that the off-chip DRAM power savings in the pipeline
are practical.

\vspace{-0.2cm}
\subsection{Comparison with State of the Art}

Table~\ref{tab:sota_full} provides a comprehensive feature comparison of SPORT against published methods across six dimensions. Among all 360\textdegree{} VR methods, SPORT is the only one that simultaneously uses WS-PSNR as both the optimization constraint and evaluation metrics. Mao~\emph{et al.}~\cite{mao2020lowlatency} do evaluate using WS-PSNR but optimize their codec rate-distortion using plain PSNR where the metric inconsistency problem comes and that SPORT formally addresses and resolved. Li~\emph{et al.}~\cite{li2019spherical}, Yu~\emph{et al.}~\cite{yu2024latitude}, and Lin~\emph{et al.}~\cite{lin2022latitude} apply latitude-aware
methods at the codec level but provide no memory-side truncation, hardware architecture, or latency result. Among memory-truncation
methods, Kim~\emph{et al.}~\cite{kim2010sbt}, Chen~\emph{et al.}~\cite{chen2018viewer}, Haidous~\emph{et al.}~\cite{haidous2022roi}, and Oswald~\emph{et al.}~\cite{oswald2024trunmem} all target 2D video with standard PSNR and report no gaze-predictive classification or DRAM power validation. The only prior method with both a fabricated chip and a latency result is Haidous~\emph{et al.}, which achieves 40\% memory saving at 11.5\,ms but their approach is evaluated on 2D content. However, SPORT-B exceeds this with 47.9\% saving at 9.33\,ms on 360\textdegree{} VR content while additionally satisfying per-region WS-PSNR thresholds and maintaining FoV losslessness (SSIM\,=\,1.000). SPORT-B is the only method in the table that satisfies all six criteria simultaneously: 360\textdegree{} VR target, WS-PSNR-consistent optimization, per-region quality guarantee, fabricated-chip validation, on-device latency within the 20\,ms VR comfort budget, and independently validated DRAM power reduction via CACTI.

\section{Conclusion}
\label{sec:conclusion}

This paper presents SPORT, a WS-PSNR-consistent bit-truncation Software--hardware co-design framework that addresses the display-path memory bottleneck in standalone VR headsets. 
Experimental results show that SPORT-B achieves ~48\% reduction in both memory power and bandwidth while maintaining lossless FoV quality (SSIM = 1.000). The adaptive SPORT-A further improves power savings to 51.6\%. Hardware validation confirms close agreement with simulations, and CACTI analysis indicates significant reductions in DRAM leakage and access energy. The system maintains a 9.33 ms latency, well within VR constraints. Overall, SPORT reduces memory power by nearly half without impacting visual quality or requiring hardware changes, and can generalize to other memory-bound video systems. Future work includes psychological subject testing and heterogeneous ASIC implementation of TrunMEM360 with DRAM for testing.

\vspace{-1.2cm}
\begin{IEEEbiographynophoto}{Md. Sajjad Hossain}
received his B.Sc. degree in Electronics and Telecommunication Engineering
from Rajshahi University of Engineering and Technology, Bangladesh in 2017, his M.S. degree
in IT Convergence Engineering from Kumoh National Institute of Technology, South Korea in 2021
and is currently pursuing a Ph.D. in Electrical Engineering from the University of Alabama.
His research interests include memory design, edge computing, IoT, and machine learning. He is the recipient of the best paper award at DCAS’26.
\end{IEEEbiographynophoto}
\vspace{-1.2cm}
\begin{IEEEbiographynophoto}{Hasibur Rahman Hemel} 
received his B.Sc. degree in Electronics and Telecommunication Engineering from Rajshahi University of Engineering and Technology, Bangladesh in 2019. He is currently pursuing a Ph.D. in Electrical Engineering from the University of Alabama. His research interests include low-power memory design, memory fault characterization, FPGA-based prototyping, hardware security, and energy-efficient hardware architecture.
\end{IEEEbiographynophoto}
\vspace{-1.2cm}
\begin{IEEEbiographynophoto} {Kyle Mooney}  (Graduate Student Member, IEEE) received his B.S. degree in computer engineering from the University of South Alabama in 2023 and is currently pursuing a Ph.D. in Electrical Engineering from the University of Alabama. His research interests include memory design, edge computing, and artificial intelligence. He is the recipient of the best paper award at DCAS’26.
\end{IEEEbiographynophoto}
\vspace{-1.2cm}
\begin{IEEEbiographynophoto} {Yiwen Xu} received his Ph.D. degree in systems and industrial engineering from the University of Arizona. He is currently an algorithmic trader and small business owner. His interests are machine learning, algorithmic trading, AI-implemented web design, and applied operations research.
\end{IEEEbiographynophoto}
\vspace{-1.3cm}

\begin{IEEEbiographynophoto} {William Oswald} Received the B.S. degree in computer engineering, a M.S. degree in electrical engineering, and a Ph.D. in systems engineering from the University of South Alabama in 2016, 2022, and 2024 respectively. He is currently employed as a senior machine learning engineer. His research interests include low-power circuits, image processing, autonomous navigation systems, physics-informed machine learning, and novel machine learning architectures. 
\end{IEEEbiographynophoto}
\vspace{-1.3cm}
\begin{IEEEbiographynophoto}{Mario Renteria-Pinon} (Member, IEEE) earned his B.S. degree in electrical and computer engineering from The University of Texas at El Paso in 2014, his M.S. degree in electrical engineering from the University of Washington in 2016, and his Ph.D. in engineering from New Mexico State University in 2023. He worked as a Postdoctoral Fellow for the IMPACT lab at the University of South Alabama, where he conducted research on AI hardware implementation, low-power memory, and privacy hardware. Currently, Dr. Renteria-Pinon is an Assistant Professor at New Mexico State University. His research interests extend to analog and mixed-signal integrated circuits, low-power sensors, data converters, and ASICs for biomedical applications.
\end{IEEEbiographynophoto}
\vspace{-1.3cm}
\begin{IEEEbiographynophoto} {Hritom Das} (Senior Member, IEEE) received the Ph.D. degree in electrical and computer engineering from North Dakota State University, Fargo, ND, USA, in 2020. He was a postdoctoral research associate with the University of Tennessee, Knoxville, where he was affiliated with TENNLab. He is currently an Assistant Professor with the Department of Electrical and Computer Engineering, Oklahoma State University, Stillwater, OK, USA. His research interests include low-power VLSI circuit design, memory devices, AR/VR, data privacy, neuromorphic system design, and optimization techniques. He is the recipient of the best paper award at DCAS’26.
\end{IEEEbiographynophoto}
\vspace{-1.3cm}

\begin{IEEEbiographynophoto} {Zhenlin Pei} (Graduate Student Member, IEEE) holds the M.S. and Ph.D. degrees in electrical engineering from Columbia University and the University of Texas at Arlington. During his Ph.D., he collaborated with IMEC to develop CACTI++, which bridges gaps in rapid EDA-based co-design and co-optimization from the transistor to the system level. He is a postdoctoral fellow in electrical \& computer engineering at the University of Alabama. He was a senior design engineer in the IP group for tapeout at Cadence for four years. Research interests include CAD/EDA, DTCO \& STCO, computing methodologies and systems integrating emerging technologies, circuits, VLSI, and AI, including edge AI and neuromorphic computing.
\end{IEEEbiographynophoto}

\vspace{-1.3cm}

\begin{IEEEbiographynophoto} {Jinhui Wang} (Senior Member, IEEE) is currently a Full Professor and Larry Drummond Endowed Chair with the Department of Electrical and Computer Engineering at the University of Alabama, Tuscaloosa, AL, USA. His research interests include: (1) VLSI System, Digital and Mixed-Signal Integrated Circuit (IC) Design, 3D and 2.5D IC Design, and Emerging Memory; (2) AI Hardware Design, Post/Beyond CMOS Device, such as Memristor-Based Neuromorphic Computing System; and (3) Post/Beyond CMOS Devices Enabled Cybersecurity and Internet of Things (IoT) Systems. He has published over 200 refereed journal/conference papers and book chapters, as well as 31 patents in the area of emerging semiconductor technologies. His previous work has received Best Paper Awards or nominations at DATE 2021, ISVLSI 2019, ISLPED 2016, ISQED 2016, and EIT 2016.
\end{IEEEbiographynophoto}

\vspace{-1.3cm}

\begin{IEEEbiographynophoto}{Na Gong (M’13)}  received the Ph.D. degree in computer science and engineering from the State University of New York, Buffalo, in 2013. Currently, Dr. Gong is a Full Professor and Larry Drummond Endowed Chair in the Department of Electrical and Computer Engineering at the University of Alabama. Her research interests include power-efficient computing circuits and systems, memory optimization, AI hardware, and hardware privacy. She is the recipient of the best paper award at DCAS’26, EIT’16, the best paper nomination from ISVLSI’19, ISQED’16, and ISLPED’16.
\end{IEEEbiographynophoto}

\vfill

\end{document}